\PassOptionsToPackage{quiet}{fontspec}
\documentclass[a4paper,12pt]{article}
\usepackage[a4paper,top=2.54cm, bottom=2.54cm, left=3.2cm, right=3.2cm]{geometry}
\usepackage{amsmath,amssymb,amsthm,bm}
\usepackage{makecell}
\usepackage{graphicx}
\usepackage{float}
\usepackage{xcolor}
\usepackage{natbib}
\usepackage[sort&compress]{gbt7714}
\usepackage{indentfirst}
\usepackage{multirow}
\usepackage{threeparttable}

\newtheorem{assumption}{Assumption}
\newtheorem{theorem}{Theorem}

\newtheorem{lemma}{Lemma}
\newtheorem{remark}{Remark}

\usepackage{authblk}
\usepackage{algorithm}
\usepackage{algpseudocode}
\usepackage{rotating}
\title{{\Large\textbf{A Principal Square Response Forward Regression Method for Dimension Reduction}}}

\author[1]{Zheng Li \thanks{First author: liz768@nenu.edu.cn}}

\author[1]{Yunhao Wang}

\author[1]{Wei Gao \thanks{Corresponding author: gaow@nenu.edu.cn}}

\author[2]{Hon Keung Tony Ng }
\affil[1]{Key Laboratory for Applied Statistics of MOE, School of Mathematics and Statistics, Northeast Normal University, Changchun 130024, China.}

\affil[2]{Department of Mathematical Sciences, Bentley University, Waltham, MA 02452, USA}

\begin{document}

\date{}

\maketitle{}

\maketitle

\begin{abstract}
Dimension reduction techniques, such as Sufficient Dimension Reduction (SDR), are indispensable for analyzing high-dimensional datasets. This paper introduces a novel SDR method named Principal Square Response Forward Regression (PSRFR) for estimating the central subspace of the response variable Y, given the vector of predictor variables $\bm{X}$. We provide a computational algorithm for implementing PSRFR and establish its consistency and asymptotic properties. Monte Carlo simulations are conducted to assess the performance, efficiency, and robustness of the proposed method. Notably, PSRFR exhibits commendable performance in scenarios where the variance of each component becomes increasingly dissimilar, particularly when the predictor variables follow an elliptical distribution. Furthermore, we illustrate and validate the effectiveness of PSRFR using a real-world dataset concerning wine quality. Our findings underscore the utility and reliability of the PSRFR method in practical applications of dimension reduction for high-dimensional data analysis.\\

\end{abstract}

\section{Introduction\label{sec:1}}

The convergence of technological advancements, the digitalization of society, the big data paradigm, computational capabilities, and the rise of machine learning and artificial intelligence has fueled the rapid growth of high-dimensional data across numerous domains such as personalized medicine \citep{zhou2021parsimonious}, computer vision \citep{reddy2020analysis}, econometrics \citep{wang2019factor,yan2022can}, and causal inference \citep{ma2019robust}. Sufficient dimension reduction (SDR), a statistical method to extract essential information from high-dimensional data while reducing the dimensionality of the data, stands out as a pivotal tool in the analysis of high-dimensional data. The primary goal of SDR is to identify linear combinations of the independent variables that capture all the relevant information about the conditional distribution of the response variable $Y$ given the vector of predictor variables $\bm{X}$, i.e., $Y \mid \bm{X}$. By reducing the dimensionality of the data while retaining the essential information, SDR enables more efficient and effective analysis of high-dimensional datasets. In many practical applications, the underlying parametric model is often unknown. In such cases, \citet{Li1991} proposed a general model that assumes an ideal scenario where the high-dimensional vector of predictor variables, $\bm{X}$, can be reconstructed from low-dimensional projections for the purpose of regressing $Y$ on $\bm{X}$:
\begin{equation}
\label{y=f(x)}
Y=g\left(\beta_{1}^\top\bm{X},\beta_{2}^\top\bm{X},\ldots,\beta_{k}^\top\bm{X},\varepsilon\right),
\end{equation}
where $Y$ is the one-dimensional response variable, $\bm{X}=(X_{1},\ldots,X_{p})^\top$ is  $\mathit{p}$-dimensional vector of predictors, $g$ is a $\mathbb{R}^{k+1}\rightarrow \mathbb{R}$ unknown link function, $\beta_{i}$'s are unknown non-random column vectors, and $\varepsilon$ is the error term which is independent of $\bm{X}$. 

Let $\text{B}=\left( \beta_{1},\dots, \beta_{k}\right) \in\mathbb{R}^{p\times k}$ ($k\leq p$) be a $p\times k$ matrix with columns $\beta_{i}$, $i=1,\ldots,k$. Then, $Y$ depends on $\bm{X}$ only through $\text{B}^\top\bm{X}$, and the purpose of SDR is to find a matrix $\text{B}$ such that 
\begin{equation}
\label{y||X}
	Y\perp \!\!\! \perp \bm{X} \mid \text{B}^\top\bm{X}, 
\end{equation}
where $\perp \!\!\! \perp$ represents independence. The space $\mathrm{Span}(\text{B})$, Spanned by these linear combinations, is often called the effective dimension reduction (EDR) space. Note that $\text{B}$ always exists because $\text{B}$ degenerates into an identity matrix when $k = p$, and it is not unique because if Eq. \eqref{y||X} is true, then $Y \perp \!\!\! \perp \bm{X} \mid \text{P}\text{B}^\top\bm{X}$ for any nonsingular matrix $\text{P}$. Thus, the identifiable parameter here is the subspace $\mathrm{Span}(\text{B})$ rather than $\text{B}$ itself. \citet{Cook1998} introduced the central dimension reduction subspace (the smallest) to address the uniqueness problem of EDR space. If a space is an EDR space and this space is contained in any EDR space, then this space is called the central dimension reduction subspace, denoted as $\boldsymbol{S}_{Y\mid \bm{X}}$. 
Considering the mean function of regression $\mathrm{E}(Y|X)$, the purpose of SDR is to find a $p \times k$ matrix $\text{B}$ such that 
\begin{equation}
\label{Y||E(y|X)}
	Y\perp \!\!\! \perp \mathrm{E}\left({Y\mid \bm{X}}\right) \mid \text{B}^\top\bm{X},
\end{equation}
where $\mathrm{Span}(\text{B})$ is called the mean dimension reduction space \citep{cook2002dimension}, which is equivalent to 
\begin{equation*}
\label{E(y|X)}
\mathrm{E}\left({Y\mid \bm{X}}\right)=\mathrm{E}\left({Y\mid \text{B}^\top\bm{X}}\right)=h\left(\beta_{1}^\top\bm{X},\ldots,\beta_{k}^\top\bm{X}\right).
\end{equation*}
Similarly, if a subspace is a mean dimension reduction space and it is contained in any mean dimension reduction space, then this subspace is called the central mean dimension reduction subspace (the smallest), denoted as $\boldsymbol{S}_{\mathrm{E}\left({Y\mid \bm{X}}\right)}$. As expressed in \cite{cook2002dimension}, $\boldsymbol{S}_{\mathrm{E}\left({Y\mid \bm{X}}\right)}\subseteq  \boldsymbol{S}_{Y\mid \bm{X}}$, that gives Eq. \eqref{Y||E(y|X)} from Eq. \eqref{y||X}. 
%{\color{blue} In this paper, we discuss $p < n$ along with the classical SDR, although high-dimensional statistics usually imply $p>>n$. }

In general, the methods to obtain the SDR estimator $\boldsymbol{S}_{Y\mid \bm{X}}$ can be classified into two categories: inverse regression and forward regression \citep{li2018sufficient}. 

For inverse regression methods, the sliced inverse regression (SIR) was first proposed by \citet{Li1991}, where ``inverse regression" refers to the conditional expectation $\mathrm{E}\left(\bm{X}\mid Y\right)$ with $\mathrm{Var}\left\{\mathrm{E}\left(\bm{X}\mid Y\right)\right\}$ contained in $\boldsymbol{S}_{Y\mid \bm{X}}$. This is achieved by assuming a linear conditional mean (LCM) for the basis matrix $\text{B}$, which serves as a fundamental assumption for numerous dimension reduction techniques.
Given that $\boldsymbol{S}_{Y\mid \bm{X}}$ remains invariant when $\text{B}$ is multiplied by any $k\times k$ full-rank matrix, this property holds for all possible unknown $\beta_{i}$ in practice. This equivalence extends to $\bm{X}$ following an elliptically symmetric distribution.

Inspired by the SIR, other “inverse regression” methods designed for estimating  $\boldsymbol{S}_{Y\mid \bm{X}}$ have been studied. These methods include the sliced average variance estimate (SAVE) based on second-order conditional moments \citep{cook1991SAVE},
% and need the constant conditional variance assumption which is satisfied by multivariate Gaussian distribution \citep{li2018sufficient}, 
parametric inverse regression  \citep{PIR2001}, canonical correlation estimator  \citep{fung2002}, contour regression  \citep{libing2005}, inverse regression estimator (IRE) \citep{cook2005}, principal fitted components \citep{cook2007}, likelihood acquired directions \citep{cook2009likelihood}, directional reduction \citep{li2007directional}, elliptically contour inverse predictors \citep{bura2015sufficient}, elliptical sliced inverse regression  \citep{chen2022high},  generalized
kernel-based inverse regression \citep{xie2020generalizedkernel}, and functional SDR estimators \citep{li2022functional}.

For forward regression methods that focus on the conditional distribution of $Y$ given $\bm{X}$, i.e., $Y\mid \bm{X}$. \citet{li1989regression} first introduced the ordinary least squares (OLS) as a dimension reduction method,  known for its intuitive nature and straightforward algorithm. However, the primary drawback of the OLS method lies in its capability to identify only one vector, and its performance suffers notably when the dimension of $\boldsymbol{S}_{\mathrm{E}\left({Y\mid \bm{X}}\right)}$ exceeds one.  \citet{li1992principal} proposed principal Hessian directions (PHD) by finding the Hessian matrix for $\mathrm{E}\left({Y\mid \bm{X}}\right)$ with the application of Stein's lemma.  
Based on utilizing an iterative approach to estimate $\boldsymbol{S}_{\mathrm{E}\left({Y\mid \bm{X}}\right)}$ with the PHD method, 
\citet{cook2002dimension} introduced the 
iterative Hessian transformation (IHT) method. This method was further studied by  \citet{cook2004determining}. Additionally, \citet{chen2018generalizedprinciple} proposed the generalized PHD (GPHD), and \citet{luo2018secondorder} proposed the adjusted PHD (APHD) methods for mixture multivariate skew elliptical distributions and non-Gaussian predictors, respectively. The IHT, GPHD, and APHD methods can be applied to a wider range of scenarios due to their less restrictive conditions compared to the PHD method. Other forward regression dimension reduction methods, such as the minimum average variance estimator (MAVE) \citep{xia2002adaptive},  sliced regression  \citep{wang2008sliced},  ensemble of minimum average variance estimators \citep{yin2011sufficient}, semiparametric dimension reduction method \citep{ma2012semiparametric}, outer-product-gradient method (OPG) \citep{xia2002adaptive,opg2007,2014Xiaadaptive,kang2022forward} and optimal SDR \citep{bura2022sufficient}, have been developed in the literature.

%Existing dimension reduction methods offer various benefits, yet they also have limitations.  For instance, the SIR method is frequently employed due to its computational efficiency, robustness, and interpretability. However, the SIR method exhibits inferior performance relative to other methods when the link function \ $\mathit{f}$ \ is symmetric about the origin. 
%The straightforward approach for dimension reduction is the OLS method, 
%Another simple method of dimension reduction is OLS, which is very intuitive with an easy algorithm, but its biggest limitation is that it can only find one vector. When the dimension of $\boldsymbol{S}_{\mathrm{E}\left({Y\mid \bm{X}}\right)}$ is greater than one, its performance is not good. 

%In response to the limitations of commonly utilized dimension reduction methods, 
This paper introduces a principal square response forward regression (PSRFR) estimator for dimension reduction. The proposed approach leverages the OLS estimator from a fresh angle, thus resolving the challenge of OLS's limited capability to recover only one dimension. 
In contrast to the PHD method, we demonstrate that the proposed PSRFR approach may be applicable under the assumption of elliptical distributions. Moreover, the PSRFR method offers greater simplicity and intuitiveness compared to the IHT method. Additionally, it can identify more central subspace directions in certain scenarios than the PHD and IHT methods.

The rest of this paper is organized as follows. In Section \ref{PropoMthd}, we propose the PSRFR estimator and derive its consistent and asymptotic normal distribution. A simulation study is conducted in Section \ref{Simula}, demonstrating that the PSRFR estimator outperforms existing methods when the variance of each component of $\bm{X}$ is significantly different under the assumption of an elliptical distribution, and it also showcases its robustness. In Section \ref{RealData}, we investigate the Wine Quality dataset through a real data analysis. Section \ref{Concluding} provides some concluding remarks for this paper.

\section{Proposed Methodology}
\label{PropoMthd}
In this section, we introduce the proposed PSRFR estimator for $\mathrm{Span}(\text{B})$, which is the smallest dimension reduction subspace we are interested in, and examine its associated theoretical properties.
Here, we assume that the structural dimensions (the rank of matrix $\text{B}$) are known, we do not attempt to determine the dimension of the central subspace in this paper. And hence, estimating $\mathrm{Span}(\text{B})$ is equivalent to estimating the direction of the basis of  $\mathrm{Span}(\text{B})$. For convenience, we consider that $\text{B}$ satisfies $\text{B}^\top\text{B}=\textup{I}_{k}$, where $\textup{I}_{k}$ is a $k$-dimensional identity matrix.

\subsection{PSRFR Estimator}

To introduce the proposed PSRFR estimator, we start with the OLS estimator under the elliptical distributions with the following assumption. 

\begin{assumption} 
\label{Ass1}
The distribution of $\bm{X}$ is an elliptical distribution with mean $\mathrm{E}(\bm{X})=0$ and variance-covariance matrix $\mathrm{Var}(\bm{X})=\Sigma_{X}$.
\end{assumption}
The following Lemma \ref{Lemma1} shows that $\mathrm{E}(Y\bm{X})$ fall in the central mean dimension reduction subspace $\boldsymbol{S}_{\mathrm{E}\left({Y\mid \bm{X}}\right)}$. 
\begin{lemma}
\label{Lemma1}
For $Y$ and $\bm{X}$ that satisfy Eq. \eqref{Y||E(y|X)}, Assumption \ref{Ass1} implies
\begin{equation}
\label{(E(YX1))}
\mathrm{E}(Y\bm{X})=\Sigma_{X}\textup{B}\Lambda,
\end{equation}
where $\Lambda=\left(\lambda_{1}, \ldots,\lambda_{k}\right)^\top$ is a constant vector. 
\end{lemma} 
\noindent 
The proof of Lemma \ref{Lemma1} is provided in the Appendix $\textup{A.1}$. \citet{Brillinger2012}, \citet{cook2002dimension} \citet{cook2004determining},  \citet{li2018sufficient}, and \citet{li1989regression} similarly provide results analogous to those in Lemma \ref{Lemma1}.

Note that the OLS estimator is a vector representation of a basis in $\boldsymbol{S}_{\mathrm{E}\left({Y\mid \bm{X}}\right)}$, which is only a precise estimator when the structure dimension is one. According to Eq. \eqref{(E(YX1))}, $\mathrm{E}(Y\bm{X})$ is a linear combination of $\{\beta_{i}\}_{i=1}^{k}$,  
which indicates that  $\mathrm{E}(Y\bm{X})$ lies in the hyperplane Spanned by $\{\beta_{i}\}_{i=1}^{k}$. To obtain a complete set of the basis of  $\boldsymbol{S}_{\mathrm{E}\left({Y\mid \bm{X}}\right)}$, it is required to determine a set of standard orthogonal basis for $\boldsymbol{S}_{\mathrm{E}\left({Y\mid \bm{X}}\right)}$. Without loss of generality, we assume that $\bm{X}$ has mean $0$ and an identity matrix as the variance-covariance matrix, and $\{(Y_{j},\bm{X}_{j})\}_{j=1}^{n}$ is a set of independent and identically distributed samples from Eq. \eqref{Y||E(y|X)}. 

The crux of the issue lies in estimating the basis of $\boldsymbol{S}_{\mathrm{E}\left({Y\mid \bm{X}}\right)}$ using $\{\bm{Z}_{j}\}_{j=1}^{n}$, where $\bm{Z}_{j}=Y_{j}\bm{X}_{j}$. This can be intuitively solved by minimizing the sum of distances from all sample points to the hyperplane spanned by the basis matrix $\text{B}$. Given that $\text{B}^\top\text{B}=\textup{I}_{k}$, the distance from $\bm{Z}_{j}$ to the hyperplane $\mathrm{Span}(\text{B})$ can be expressed as follows:
%Now, the key to the problem is to estimate the basis of $\boldsymbol{S}_{E\left({Y\mid \bm{X}}\right)}$ by $\{\bm{Z}_{j}\}_{j=1}^{n}$, where $\bm{Z}_{j}=Y_{j}\bm{X}_{j}$, which can be solved intuitively by finding the minimum sum of distances from all sample points to the hyperplane \mathrm{Span}ned by the basis matrix $\text{B}$. Since $\text{B}^\top\text{B}=I_{k}$, the distance from $\bm{Z}_{j}$ to the hyperplane can be represented as follows: 
\begin{align*}
d_{j} =\left \| \bm{Z}_{j}-\text{B}\text{B}^\top\bm{Z}_{j} \right \|_{2}^{2} 
= \bm{Z}_{j}^\top\bm{Z}_{j}-\bm{Z}_{j}^\top\text{B}\text{B}^\top\bm{Z}_{j},
\end{align*}
where $||\cdot||_2$ represents $\mathrm{L}_{2}$ norm. Then, the minimization problem to obtain the estimator for $\mathrm{Span}(\text{B})$ can be expressed as
\begin{align*}   
\min_{B}\sum_{j=1}^{n} d_j 
&=
\min_{B}\sum_{j=1}^{n}(\bm{Z}_{j}^\top \bm{Z}_{j}-\bm{Z}_{j}^\top BB^\top \bm{Z}_{j})
\\&=\min_{\{\beta_{i}\}_{i=1}^{k}}\sum_{j=1}^{n}\left (\bm{Z}_{j}^\top \bm{Z}_{j}- \sum_{i=1}^{k}\bm{Z}_{j}^\top \beta_{i}\beta_{i}^\top \bm{Z}_{j}\right ),
\end{align*}
which is equivalent to the maximization problem
\begin{align*}
\max_{\{\beta_{i}\}_{i=1}^{k}}\sum_{j=1}^{n}\sum_{i=1}^{k}\bm{Z}_{j}^\top \beta_{i}\beta_{i}^\top \bm{Z}_{j}	&=\max_{\{\beta_{i}\}_{i=1}^{k}}\sum_{j=1}^{n}\sum_{i=1}^{k}\beta_{i}^\top \bm{Z}_{j}\bm{Z}_{j}^\top \beta_{i}
\\&=\max_{\{\beta_{i}\}_{i=1}^{k}}\sum_{i=1}^{k}\beta_{i}^\top \left ( \sum_{j=1}^{n}\bm{Z}_{j}\bm{Z}_{j}^\top  \right )\beta_{i}.
\end{align*}
It can be shown that the solution of the above optimization problem is the first $k$ eigenvectors corresponding to the first $k$ eigenvalues of $\sum_{j=1}^{n}\bm{Z}_{j}\bm{Z}_{j}^\top$. 

For the practical situation that the mean $\mathrm{E}(\bm{X})$ and variance-covariance matrix of $\bm{X}$, $\Sigma_{X}$, are unknown, 
the sample mean $\Bar{\bm{X}}$ and the sample variance $S_{n}$ based on the observed sample can be used to approximate $\mathrm{E}(\bm{X})$ and $\Sigma_{X}$, respectively. Then, the data can be transformed as $\bm{Z}_{j}=S_{n}^{-1}Y_{j}(\bm{X}_{j}-\bar{\bm{X}})$. 

The aforementioned results describe how the PSRFR estimator of $\mathrm{Span}(\text{B})$ can be obtained. Hence, a set of standard orthogonal bases in $\mathrm{Span}(\text{B})$ can also be estimated similarly. The algorithm to obtain the PSRFR estimator, namely the PSRFR algorithm, is presented in Algorithm 1.

\begin{algorithm}[H]
\caption{The PSRFR Algorithm}
\textbf{1}. Center $\bm{X}_{j}$, i.e.,  $\tilde{\bm{X}}_{j}=\bm{X}_{j}-n^{-1}\sum_{j=1}^{n}\bm{X}_{j}$.

\textbf{2}. Compute $\bm{Z}_{j}=S_{n}^{-1}Y_{j}\tilde{\bm{X}}_{j}$, where $S_{n}=(n-1)^{-1}\sum_{j=1}^{n}\tilde{\bm{X}}_{j}\tilde{\bm{X}}_{j}^\top$.

\textbf{3}. Obtain the eigendecomposition of   
$\hat{\mathcal{Z}}=n^{-1}\sum_{j=1}^{n}\bm{Z}_{j}\bm{Z}_{j}^\top$, and extract the first $k$ eigenvectors corresponding to the $k$ largest eigenvalues of $\hat{\mathcal{Z}}$ denoted by $\hat{\text{B}}=(\hat{\beta}_{i}, \ldots, \hat{\beta}_{k})$. 
\end{algorithm}

At the population level, the PSRFR estimator is based on the $p \times p$ positive-defined matrix
\begin{equation}
\label{PSRFR}
\Sigma_{X}^{-1}\mathrm{E}\left[Y^{2}\{\bm{X}-\mathrm{E}(\bm{X})\}\{\bm{X}-\mathrm{E}(\bm{X})\}^\top\right]\Sigma_{X}^{-1}, 
\end{equation}
which is associated with the principal components analysis (PCA) and the PHD method, as described below.

For PCA, the principal components of $\bm{X}$ are defined as the set of linear combinations of $\bm{X}$ that have the largest variances. The first principal component of $\bm{X}$ can be obtained by solving the maximization problem
\begin{equation*}
\max_{||\gamma ||_{2}=1} \gamma^\top \mathrm{E}\left[\{\bm{X}-\mathrm{E}(\bm{X})\}\{\bm{X}-\mathrm{E}(\bm{X})\}^\top\right]\gamma,
\end{equation*}
in which the solution is the first eigenvector of $\mathrm{E}\left[\{\bm{X}-\mathrm{E}(\bm{X})\}\{\bm{X}-\mathrm{E}(\bm{X})\}^\top\right]$. Similarly, the first $k$ principal components of $\bm{X}$ are the first $k$ eigenvectors of $\Sigma_{X}$. After obtaining the principal components, a regression model such as the one presented in Eq. \eqref{y=f(x)} can be constructed. 
However, projecting the $p$-dimensional predictor variables onto lower dimensions first might inadvertently result in an inaccurate relationship between $Y$ and the original $\bm{X}$. In contrast, the proposed PSRFR method systematically establishes the connections between $\bm{X}$ and $Y$ at the beginning of the dimension reduction process.

%However, the working of first projecting the $p$-dimensional predictor variables onto the low-dimensions might lead to finding a wrong relationship between $Y$ and the original $\bm{X}$. To this end, PSRFR grasps the links between $\bm{X}$ and $Y$ at the beginning. 

For the PHD method, it is based on the Hessian matrix of twice differentiable regression function $\mathrm{E}(Y\mid \bm{X})$, denoted as $H_{m}(\bm{X})$. By the chain rule, we have  
\begin{equation*}
H_{m}(\bm{X}) =\frac{\partial^{2} \mathrm{E}(Y\mid \bm{X})}{\partial \bm{X}\partial \bm{X}^\top}=\text{B} \dfrac{\partial^{2} \mathrm{E}(Y\mid \text{B}^\top\bm{X})}{\partial (\text{B}^\top\bm{X})\partial (\bm{X}^\top\text{B})}\text{B}^\top.
\end{equation*}
Let $\overline{H_m}(\bm{X})=\mathrm{E}\{H_{m}(\bm{X})\}$. If $\bm{X}$ follows a normal distribution,  $\overline{H_m}(\bm{X})$ can be transformed into an easily solvable form by Stein's lemma, $\Sigma_{X}^{-1}\Sigma_{YXX}\Sigma_{X}^{-1}$,
where 
\begin{equation}
\label{PHD}
    \Sigma_{YXX}=\mathrm{E}\left[\{Y-\mathrm{E}(Y)\}\{\bm{X}-\mathrm{E}(\bm{X})\}\{\bm{X}-\mathrm{E}(\bm{X})\}^\top\right],
\end{equation}
which is known as the response-based ($\mathrm{y}$-based) version of the PHD method. In contrast to the PHD approach, besides extending the normal distribution assumption to the elliptical distribution assumption, the proposed PSRFR replaces $Y$ with $Y^{2}$ in Eq. \eqref{PHD}, potentially leading to notable effects on the estimators.
%Compared to the PHD approach, in addition to generalizing the normal distribution assumption to the elliptical distribution assumption, the proposed PSRFR formally replaces $Y$ by $Y^{2}$ in Eq. \eqref{PHD}, which might have a significant impact on the estimators.

\begin{remark}
\label{cmpPHD}
    The major difference between the proposed PSRFR approach and the  PHD approach is the term  $\mathrm{E}(Y^{2}\mid\bm{X})$ and $\mathrm{E}(Y\mid\bm{X})$ involved according to the law of iterated expectation. Consider the following model in \citet{li2018sufficient}: \begin{equation*}
    Y=f\left(\beta_{1}^\top\bm{X}\right)+g\left(\beta_{2}^\top\bm{X}\right)\varepsilon,
    \end{equation*}
    where $\varepsilon$ is dependent on $\bm{X}$ with zero mean,  $\beta_{1},\beta_{2} \in \mathbb{R}^{p}$, and $f$ and $g$ are unknown link functions. 
    Since $\mathrm{E}(Y\mid\bm{X})=f(\beta_{1}^\top\bm{X})$ and  $\mathrm{E}(Y^{2}\mid\bm{X})=f^{2}(\beta_{1}^\top\bm{X})+g^{2}(\beta_{2}^\top\bm{X})\mathrm{Var}(\varepsilon)$, the PSRFR method can identify more central subspace directions than the PHD method. Moreover, 
    \begin{equation*}
        \mathrm{E}\left(Y^2\mid \bm{X}\right)=\mathrm{Var}(Y\mid \bm{X})+\mathrm{E}(Y\mid\bm{X})\mathrm{E}(Y\mid\bm{X})
    \end{equation*}
    according to the definition of conditional variance. Here, $\mathrm{E}(Y^2\mid \bm{X})$ contains $\mathrm{Var}(Y\mid \bm{X})$, and not just $\mathrm{E}(Y\mid\bm{X})$.

\end{remark}
 Remark \ref{cmpPHD} shows the PSRFR might identify more central subspace directions in the above example under the model in Eq. \eqref{y=f(x)}; hence, the EDR space estimated by the PSRFR is no longer limited to the central mean subspace. Furthermore, we are interested in  
\begin{equation}
\label{E(Y^2|X)}
    \mathrm{E}\left(Y^{2}\mid \bm{X}\right)=\mathrm{E}\left(Y^{2}\mid \text{B}^\top\bm{X}\right)=H\left(\beta_{1}^\top\bm{X}, \ldots,\beta_{k}^\top\bm{X}\right).
\end{equation}
%{\color{red} To avoid misuse of notations, we use the same notation $k$ for structure dimension and $\text{B}$ for basis matrix in Eq. \eqref{E(Y^2|X)}. They might be different from Eq. \eqref{(E(YX1))} in some cases.}
%,that is, the central subspace directions obtained from Eq. \eqref{E(Y^2|X)} is more than \eqref{(E(YX1))}.
Due to the non-uniqueness of $\text{B}$ and for ease of expression in the remainder of the paper, $\mathrm{Span}(\text{B})$ is used to represent the dimension reduction subspace in Eq. \eqref{E(Y^2|X)}, where $\mathrm{Span}(\text{B}) \subseteq \boldsymbol{S}_{Y\mid \bm{X}}$, which makes it easy to derive Eq. \eqref{E(Y^2|X)} from Eq. \eqref{y=f(x)}.

\subsection{Asymptotic Properties}

In this subsection, we will prove that $\{\hat{\beta}_{i}\}_{i=1}^{k}$ converge to a set of standard orthogonal basis for $\mathrm{Span}(\text{B})$ under some mild conditions. 
\begin{assumption}
\label{Ass2}
Under the model in Eq. \eqref{E(Y^2|X)}, the inequality 
\begin{equation}
\label{eq9}
   \mathrm{E} \left\{Y^{2}  \left(\beta^\top\Sigma_{X}^{-1}\bm{X}\right)^{2}\right\}>\mathrm{E}\left\{Y^{2}\left(\alpha^\top\Sigma_{X}^{-1}\bm{X}\right)^{2} \right\}
\end{equation}
holds, where $\beta \in \mathrm{Span}(\text{B})$, $\alpha \in \mathrm{Span}(\text{B})^{\perp}$, $\left \| \beta \right \|_{2}=\left \| \alpha \right \|_{2}=1$, and the symbol $\perp$ represents the orthogonal complement.
\end{assumption} 
In Assumption \ref{Ass2}, the term on the left-hand side of the inequality in Eq. \eqref{eq9} 
\begin{equation*}
\mathrm{E} \left\{Y^{2}  \left(\beta^\top\Sigma_{X}^{-1}\bm{X}\right)^{2}\right\}=\left(\beta^\top\text{B}\Lambda\right)^2+\mathrm{Var}\left(Y\beta^\top\Sigma_{X}^{-1}\bm{X}\right),
\end{equation*} 
which represents the sum of a fixed term $\left(\beta^\top\text{B}\Lambda\right)^2$  %{\color{red} ($\text{B}\Lambda$ has the same meaning as Eq. \eqref{(E(YX1))})} 
and the variance of some random variables, 
and the term of the right-hand side of the inequality in Eq. \eqref{eq9} 
\begin{equation*}
\mathrm{E}\left\{Y^{2}\left(\alpha^\top\Sigma_{X}^{-1}\bm{X}\right)^{2}\right\}=\mathrm{Var}\left(Y\alpha^\top\Sigma_{X}^{-1}\bm{X}\right)
\end{equation*}
represents the variance of some random variables.  

\begin{theorem}
\label{thm1}
For Model in Eq. \eqref{E(Y^2|X)}, under Assumptions \ref{Ass1} and \ref{Ass2}, the first $k$ eigenvectors corresponding to the first $k$ eigenvalues of~ $\mathrm{E}(\bm{Z}\bm{Z}^\top)$ are the basis of $\mathrm{Span}(\textup{B})$, where $\bm{Z}=\Sigma_{X}^{-1}Y\bm{X}$, and $\mathrm{E}(\bm{Z}\bm{Z}^\top)-\Sigma_{X}^{-1}\mathrm{E}\left(\textup{G}\right)$ is contained in $\mathrm{Span}(\textup{B})$, where $\mathrm{E}\left(\textup{G}\right)=\mathrm{E}( Y^{2} w_{k+1}^{2})$, $w_{k+1}$ is the $(k+1)$-th element of  
  $\bm{W}=\textup{C}\bm{X}=(w_{1},\dots,w_{k},w_{k+1},\dots,w_{p})^\top$, and $\textup{C}=(\beta_{1},\dots,\beta_{k},\alpha_{k+1},\dots,\alpha_{p})^\top\equiv \left( \textup{B},\textup{A}\right)^\top$ is an orthogonal matrix.
\end{theorem}
The proof of Theorem \ref{thm1} is provided in Appendix $\textup{A.2}$. 

Although $\mathrm{E}(\bm{Z}\bm{Z}^\top)$ does not fall completely in $\mathrm{Span}(\text{B})$, a set of basis can still be found through the eigendecomposition of $\mathrm{E}(\bm{Z}\bm{Z}^\top)$, namely, 
\begin{equation}
\label{psi1}
\mathrm{E}\left(\bm{Z}\bm{Z}^\top\right)=\begin{pmatrix}
 \textup{Q}_{1}& \textup{Q}_{0}
\end{pmatrix}\begin{pmatrix}
\Psi_{1} & 0\\ 
0 & \Psi_{0}
\end{pmatrix}\begin{pmatrix}
\textup{Q}_{1}^\top\\ 
\textup{Q}_{0}^\top
\end{pmatrix}=\textup{Q}_{1}\Psi_{1}\textup{Q}_{1}^\top+\textup{Q}_{0}\Psi_{0}\textup{Q}_{0}^\top,
\end{equation}
where $\textup{Q}=\left ( \textup{Q}_{1},\textup{Q}_{0} \right )$ is a $p$-dimensional orthogonal matrix, $\Psi_{1}$ and $\Psi_{0}$ are diagonal matrices with dimensions $k$ and $p-k$, respectively. The diagonal elements in $\Psi_{1}$ and $\Psi_{0}$ are ordered from large to small.  
In the proof of Theorem \ref{thm1}, we show that 
\begin{equation}
\label{gamma1}
\mathrm{E}\left(\bm{Z}\bm{Z}^\top\right)=\text{B}\Gamma _{1}\text{B}^\top+\text{A}\Gamma _{2}\text{A}^\top,
\end{equation}
where $\Gamma_{1}$ is a $k \times k$ positive-defined matrix with the ($i, j$)-th elements is \linebreak $\mathrm{E}\{ H(w_{1},\dots,w_{k}) w_{i}w_{j}\}$, $\Gamma _{2}$ is a $(p-k) \times (p-k)$ diagonal matrix with diagonal elements $\mathrm{E}(\text{G})$.
 Consider that $\Gamma_1$ is not a diagonal matrix in Eq. \eqref{gamma1}, we can rewrite $\text{B}\Gamma _{1}\text{B}^\top=\text{B}\text{V}\Phi \text{V}^\top\text{B}^\top$ by eigendecomposition, where $\text{V}$ and $\Phi$ are $k$-dimensional orthonormal and diagonal matrices, respectively. 
Then, it is sufficient to show $\mathrm{Span}(\text{B})=\mathrm{Span}(\text{B}\text{V})=\mathrm{Span}(\text{Q}_{1})$, where the pivotal challenge arises in identifying $\mathrm{Span}(\text{B})$ through eigendecomposition of $\mathrm{E}(\bm{Z}\bm{Z}^\top)$. 
Under Assumption \ref{Ass2}, the eigenvalues corresponding to the eigenvectors of $\Gamma_{1}$ exceed those corresponding to the eigenvectors of $\Gamma_{2}$, which guarantees that the first $k$ eigenvectors of $\mathrm{E}(\bm{Z}\bm{Z}^\top)$ corresponding to the first $k$ eigenvalues must be the basis of $\mathrm{Span}(\text{B})$.
Thus, the basis can be determined by finding the first $k$ eigenvectors corresponding to the first $k$ eigenvalues. 

\begin{remark}
\label{normalrmk}
 For the model in Eq. \eqref{E(Y^2|X)}, if $\bm{X}$ follows a multivariate Gaussian distribution, then  $\Sigma_{X}^{-1}\mathrm{E}[\{Y^{2}-\mathrm{E}(Y^{2})\}\bm{X}\bm{X}^\top]\Sigma_{X}^{-1}$ is contained in $\mathrm{Span}(\text{B})$.
In the case of Remark \ref{cmpPHD}, the proposed PSRFR method can still identify more central subspace directions. More details are provided in the proof of Theorem \ref{thm1} presented in Appendix $\textup{A.2}$. 
Moreover, based on the idea that square response is contained in the PSRFR method, if one first transforms the response variable $Y$ into $Y^2$ and then applies those SDR methods that focus on the $\mathrm{E}(Y\mid \bm{X})$, 
more central subspace directions can be identified in the case of Remark \ref{cmpPHD}. 
\end{remark}

Following Theorem \ref{thm1}, the asymptotic properties of the estimator $\{\hat{\beta}_{i}\}_{i=1}^{k}$ can be established, and the results are given in Theorem \ref{Thm2}.

\begin{theorem}
\label{Thm2}
For the model in Eq. \eqref{E(Y^2|X)}, under Assumptions \ref{Ass1} and \ref{Ass2}, if ~$\mathrm{E}\left\{\mathrm{E}(Y\mid \bm{X})^{2}\right\}$\\$<\infty$ holds, then
\begin{equation}
\label{F1}
	\mathrm{Span}(\hat{\beta}_{1},\ldots,\hat{\beta}_{k})\stackrel{\mathrm{Pr}}{\longrightarrow}\mathrm{Span}(\beta_{1},\ldots,\beta_{k}), 
\end{equation}
and $\mathrm{Span}(\hat{\textup{B}}$) converges to $\mathrm{Span}(\textup{B})$ at rate $n^{1/2}$. 
In addition, if ~$\mathrm{Var}\left \{ \mathrm{vec}(\bm{Z}\bm{Z}^\top) \right\}$ exists, then 
\begin{equation}
    \sqrt{n}\left [ \mathrm{vec}(\hat{\mathcal{Z}})- \mathrm{vec}\left\{\mathrm{E}\left(\bm{Z}\bm{Z}^\top\right)\right\} \right ]\stackrel{\mathrm{L}}{\longrightarrow} \mathrm{N}\left[ 0, \mathrm{Var}\left \{ \mathrm{vec}\left(\bm{Z}\bm{Z}^\top\right) \right \} \right], 
\end{equation}
 where $\mathrm{vec}(\cdot)$ is the operator that maps a symmetric matrix to a vector by stacking the 
main diagonal and the elements below the main diagonal by columns, i.e., if $S$ is a $p \times p$ symmetric matrix 
\begin{eqnarray*}
S = 
\begin{pmatrix} 
s_{11} & s_{12} & s_{13} & \cdots & s_{1p} \\
s_{12} & s_{22} & s_{23} & \cdots & s_{2p} \\
s_{13} & s_{23} & s_{33} & \cdots & s_{3p} \\
\vdots & \vdots & \vdots & \cdots & \vdots \\ 
s_{1p} & s_{2p} & s_{3p} & \cdots & s_{pp} \\
\end{pmatrix}, 
\end{eqnarray*}
then $\mathrm{vec}(S) = (s_{11}, \ldots, s_{1p}, s_{22},\ldots, s_{2p}, s_{33}, \ldots, s_{3p}, \ldots, s_{pp})^\top$. 
%: if $S$ is a matrix with columns $s_{1},\dots,s_{p}$, then $\mathrm{vec}(S)=\left(s_{1}^\top,\dots,s_{p}^\top\right)^\top$.

\end{theorem}
The proof of Theorem \ref{Thm2} is provided in Appendix $\textup{A.3}$. 
Note that 
\begin{align*}
    \mathrm{E}\left(Y^{2}\right)=\mathrm{E}\left\{\mathrm{E}\left(Y^{2}\mid \bm{X}\right)\right\}=\mathrm{E}\left\{H\left(\beta_{1}^\top\bm{X},\cdots,\beta_{k}^\top\bm{X}\right)\right\} \geq \mathrm{E}\left\{\mathrm{E}\left(Y\mid \bm{X}\right)^{2}\right\},
\end{align*}
hence, in Theorem \ref{Thm2}, instead of the condition $\mathrm{E}\{\mathrm{E}(Y\mid \bm{X})^{2}\}<\infty$, we can use $\mathrm{E}(Y^{2})<\infty$.  
Because $\hat{\mathcal{Z}}$, $\bm{Z}\bm{Z}^\top $, and $\mathrm{E}(\bm{Z}\bm{Z}^\top)$ are $p \times p$ symmetric matrices, the dimensions of $\mathrm{vec}(\hat{\mathcal{Z}})$, $\mathrm{vec}(\bm{Z}\bm{Z}^\top )$, and $\mathrm{vec}\{\mathrm{E}(\bm{Z}\bm{Z}^\top)\}$ are $p \times (p+1)/2$. 
Notice that $\mathrm{Var}\left \{ \mathrm{vec}(\bm{Z}\bm{Z}^\top) \right \}$ is a $p(p+1)/2$ by $p(p+1)/2$ matrix. We represent the $\mathrm{Var}\left \{ \mathrm{vec}(\bm{Z}\bm{Z}^\top) \right \}$ by the Kronecker product $\otimes  $ in the proof of Theorem \ref{Thm2},  the dimension of $(\bm{Z}\bm{Z}^\top)\otimes(\bm{Z}\bm{Z}^\top)$ is $p^2$ by $p^2$. Although the Kronecker product representation has a degenerate variance (all the elements in symmetric matrices $\bm{Z}\bm{Z}^\top$ are used), this representation is for notation convenience only as our concern is the convergence of each element of variance of the random matrix $\bm{Z}\bm{Z}^\top$.

Theorem \ref{Thm2} shows that  $\mathrm{Span}(\hat{\text{B}}$) are $\sqrt{n}$-consistency estimators of $\mathrm{Span}(\text{B})$ by the law of large number. Although the first $k$ eigenvectors corresponding to the first $k$ eigenvalue of $\mathrm{E}(\bm{Z}\bm{Z}^\top)$ are not a set of original basis in Eq. \eqref{E(Y^2|X)}, they represent the same space. Moreover, $\hat{\mathcal{Z}}=n^{-1}\sum_{j=1}^{n}(\bm{Z}_{j}\bm{Z}_{j}^\top)$ is a $\sqrt{n}$-consistency estimator of $\mathrm{E}(\bm{Z}\bm{Z}^\top)$. 
Additionally, if $\mathrm{Var}\left \{ \mathrm{vec}(\bm{Z}\bm{Z}^\top) \right \}$ exists, the asymptotic normality property is obtained by the central limit theorem. 

\section{Monte Carlo Simulation Study}
\label{Simula}
In this section, we evaluate the performance of the proposed PSRFR method by using a Monte Carlo simulation study. To measure the distance between the true subspace $\mathrm{Span}(\text{B})$ and the corresponding estimator $\mathrm{Span}(\hat{\text{B}})$ for $\hat{\text{B}}=(\hat{\beta_{1}}, \ldots,\hat{\beta_{k}})$, we consider the trace correlation defined as  \citep{ferre1998determining,dong2015robust} 
\begin{equation}
\label{eq14}
R   = \dfrac{\mathrm{trace}\left (P_{\text{B}}P_{\hat{\text{B}}}\right )}{k},
\end{equation}
where $P_{\text{B}}=\text{B}(\text{B}^\top\text{B})^{-1}\text{B}^\top$ denotes the projection matrix. 
Without loss of generality, we assume $\hat{\text{B}}$ is a column-orthogonal matrix due to the property of $\text{B}$. Otherwise, the Gram-Schmidt ortho-normalization method can be used and will not change the subspace. Then, the trace correlation based on the estimator $\hat{\text{B}}$ in Eq. \eqref{eq14} can be calculated as
\begin{equation}
R  =\dfrac{\mathrm{trace}\left (\hat{\text{B}}^\top \text{B}\text{B}^\top\hat{\text{B}}\right )}{k}.
\end{equation}
Here, the trace correlation $R$ can be used to evaluate and compare the performance of different estimation methods. The trace correlation is a value between 0 and 1, and a larger value of $R$ indicates a better estimator $\hat{\text{B}}$.  In the following subsections, we consider $\bm{X}$ follows elliptical distributions in Section \ref{sec3.1} to investigate the validity of the proposed PSRFR method, and $\bm{X}$ follows non-elliptical distributions in Section \ref{sec3.2} to evaluate the robustness of different methods. 

\subsection{Elliptical distributions}
\label{sec3.1} 

In this subsection, two types of elliptical distributions,
normal and non-normal distributions are considered as the distribution of the predictor variables $\bm{X}$.

\subsubsection{Normal distribution}
\label{sec:311}

First, we compare the proposed PSRFR method to the PHD and IHT methods under the model in \cite{li2018sufficient} described in Remark \ref{cmpPHD}. The following two models are considered in the simulation study: 
\begin{itemize} 
\item {\bf Model [N1]}
\begin{equation*}
Y=\beta_{1}^\top\bm{X}+\beta_{2}^\top\bm{X}\cdot\varepsilon.
\end{equation*}
\item {\bf Model [N2]}
\begin{equation*}
Y=\sin(\beta_{1}^\top\bm{X})+(|\beta_{2}^\top\bm{X}+1|)^{1/2}\cdot\varepsilon.
\end{equation*} 
\end{itemize} 
We consider $p=\mathrm{dim}(\bm{X}) = 10$, $\bm{X} \sim \mathrm{N}_{10}(0, \Sigma_{norm})$,  $\varepsilon\sim \mathrm{N}(0,1)$,  $\beta_{1}=(1,0,0,\dots,0)$ and $\beta_{2}=(0,1,0,\dots,0)$, where $\Sigma_{norm}$ is a diagonal matrix with the diagonal elements $(1,2,3, \ldots, 10)$, and the sample sizes are $n = 100$, $300$ and $500$.

We use the trace correlation as a comparison criterion and also consider the cosine similarity criteria. 
Specifically, the cosine similarity for $\beta_{1}$ is defined as  
\begin{equation*}
|\cos_1|=\max\{|\hat{\beta}_{1}^\top\beta_{1}|/\| \hat{\beta}_{1}^\top\|, |\hat{\beta}_{1}^\top\beta_{2}|/\| \hat{\beta}_{1}^\top\|\},
\end{equation*}
which is the absolute value of the cosine of the closest true direction to $\hat{\beta}_{1}^\top$. Similarly, the cosine similarity for $\beta_{2}$ is defined as  
\begin{equation*}
|\cos_2|=\max\{|\hat{\beta}_{2}^\top\beta_{2}|/\| \hat{\beta}_{2}^\top\|, |\hat{\beta}_{2}^\top\beta_{1}|/\| \hat{\beta}_{2}^\top\|\}.
\end{equation*}
Larger values of $|\cos_1|$ and $|\cos_2|$ indicate a better estimator $\hat{\text{B}}$. 

 The computer program written in R \citep{Rsoft2024} for the implementation of the proposed PSRFR method is provided in Appendix $\textup{A.4}$. The PHD and IHT methods are implemented in the R packages {\tt dr}  \citep{drpackage2002} and {\tt itdr} \citep{itdrpackage2021}, respectively.
The averages and standard deviations (SDs) of 
$R$, $|\cos_1|$ and $|\cos_2|$ for the proposed PSRFR method, the PHD method, and the IHT method based on 1000 simulations are reported in Table \ref{comparetable}. 

 %{\clr The PHD  is available in the “dr” package in R.}Considering the following two models:

\begin{table}[!ht]
\centering
\caption{Averages and standard deviations of the trace correlation and cosine similarities for the PSRFR, PHD, and IHT methods for models {\bf [N1]} and {\bf [N2]} based on 1000 simulations with different sample sizes.}
\label{comparetable}
    \setlength{\tabcolsep}{3pt} % Default value: 6pt
    \renewcommand{\arraystretch}{1.2}
{\footnotesize
\begin{tabular}{ccccccccccc}
\hline
\multicolumn{1}{c}{} &  & \multicolumn{9}{c}{{\bf Model [N1]}}  \\ \hline &  & \multicolumn{3}{c}{$n=100$}   & \multicolumn{3}{c}{$n=300$}   & \multicolumn{3}{c}{$n=500$} \\ 
Method&   & $R$ & $|\cos_1|$ & $|\cos_2|$& $R$  & $|\cos_1|$  & $|\cos_2|$  & $R$  & $|\cos_1|$  & $|\cos_2|$ \\ \hline
PSRFR   & \begin{tabular}[c]{@{}c@{}}Average\\ SD\end{tabular} & \begin{tabular}[c]{@{}c@{}}0.917\\ 0.053\end{tabular} & \begin{tabular}[c]{@{}c@{}}0.833\\ 0.211\end{tabular} & \begin{tabular}[c]{@{}c@{}}0.795\\ 0.259\end{tabular} & \begin{tabular}[c]{@{}c@{}}0.970\\ 0.016\end{tabular} & \begin{tabular}[c]{@{}c@{}}0.905\\ 0.152\end{tabular} & \begin{tabular}[c]{@{}c@{}}0.893\\ 0.230\end{tabular} & \begin{tabular}[c]{@{}c@{}}0.981\\ 0.011\end{tabular} & \begin{tabular}[c]{@{}c@{}}0.942\\ 0.101\end{tabular} & \begin{tabular}[c]{@{}c@{}}0.934\\ 0.188\end{tabular} \\ 
PHD                         & \begin{tabular}[c]{@{}c@{}}Average\\ SD\end{tabular} & \begin{tabular}[c]{@{}c@{}}0.569\\ 0.157\end{tabular} & \begin{tabular}[c]{@{}c@{}}0.611\\ 0.232\end{tabular} & \begin{tabular}[c]{@{}c@{}}0.433\\ 0.244\end{tabular} & \begin{tabular}[c]{@{}c@{}}0.602\\ 0.137\end{tabular} & \begin{tabular}[c]{@{}c@{}}0.665\\ 0.217\end{tabular} & \begin{tabular}[c]{@{}c@{}}0.432\\ 0.233\end{tabular} & \begin{tabular}[c]{@{}c@{}}0.607\\ 0.137\end{tabular} & \begin{tabular}[c]{@{}c@{}}0.688\\ 0.206\end{tabular} & \begin{tabular}[c]{@{}c@{}}0.421\\ 0.228\end{tabular} \\ 
IHT                         & \begin{tabular}[c]{@{}c@{}}Average\\ SD\end{tabular} & \begin{tabular}[c]{@{}c@{}}0.560\\ 0.103\end{tabular} & \begin{tabular}[c]{@{}c@{}}0.885\\ 0.062\end{tabular} & \begin{tabular}[c]{@{}c@{}}0.332\\ 0.129\end{tabular} & \begin{tabular}[c]{@{}c@{}}0.604\\ 0.103\end{tabular} & \begin{tabular}[c]{@{}c@{}}0.958\\ 0.021\end{tabular} & \begin{tabular}[c]{@{}c@{}}0.137\\ 0.081\end{tabular} & \begin{tabular}[c]{@{}c@{}}0.603\\ 0.100\end{tabular} & \begin{tabular}[c]{@{}c@{}}0.974\\ 0.013\end{tabular} & \begin{tabular}[c]{@{}c@{}}0.376\\ 0.065\end{tabular} \\ \hline
\multicolumn{1}{l}{}  & \multicolumn{1}{l}{}   & \multicolumn{9}{c}{{\bf Model  [N2]}}\\ 
\hline &  & \multicolumn{3}{c}{$n=100$} & \multicolumn{3}{c}{$n=300$}  & \multicolumn{3}{c}{$n=500$}   \\ 
Method&  &$ R$ & $|\cos_1|$  & $|\cos_2|$ & $R$ & $|\cos_1|$  & $|\cos_2|$  & $R$  & $|\cos_1|$& $|\cos_2|$\\ \hline
PSRFR                         & \begin{tabular}[c]{@{}c@{}}Average\\ SD\end{tabular} & \begin{tabular}[c]{@{}c@{}}0.866\\ 0.088\end{tabular} & \begin{tabular}[c]{@{}c@{}}0.852\\ 0.189\end{tabular} & \begin{tabular}[c]{@{}c@{}}0.777\\ 0.242\end{tabular} & \begin{tabular}[c]{@{}c@{}}0.958\\ 0.026\end{tabular} & \begin{tabular}[c]{@{}c@{}}0.941\\ 0.095\end{tabular} & \begin{tabular}[c]{@{}c@{}}0.921\\ 0.183\end{tabular} & \begin{tabular}[c]{@{}c@{}}0.974\\ 0.013\end{tabular} & \begin{tabular}[c]{@{}c@{}}0.969\\ 0.048\end{tabular} & \begin{tabular}[c]{@{}c@{}}0.958\\ 0.129\end{tabular} \\ 
PHD                         & \begin{tabular}[c]{@{}c@{}}Average\\ SD\end{tabular} & \begin{tabular}[c]{@{}c@{}}0.520\\ 0.165\end{tabular} & \begin{tabular}[c]{@{}c@{}}0.469\\ 0.273\end{tabular} & \begin{tabular}[c]{@{}c@{}}0.381\\ 0.251\end{tabular} & \begin{tabular}[c]{@{}c@{}}0.539\\ 0.163\end{tabular} & \begin{tabular}[c]{@{}c@{}}0.495\\ 0.269\end{tabular} & \begin{tabular}[c]{@{}c@{}}0.394\\ 0.257\end{tabular} & \begin{tabular}[c]{@{}c@{}}0.541\\ 0.164\end{tabular} & \begin{tabular}[c]{@{}c@{}}0.475\\ 0.263\end{tabular} & \begin{tabular}[c]{@{}c@{}}0.401\\ 0.255\end{tabular} \\ 
IHT                         & \begin{tabular}[c]{@{}c@{}}Average\\ SD\end{tabular} & \begin{tabular}[c]{@{}c@{}}0.575\\ 0.084\end{tabular} & \begin{tabular}[c]{@{}c@{}}0.828\\ 0.084\end{tabular} & \begin{tabular}[c]{@{}c@{}}0.269\\ 0.130\end{tabular} & \begin{tabular}[c]{@{}c@{}}0.571\\ 0.079\end{tabular} & \begin{tabular}[c]{@{}c@{}}0.934\\ 0.031\end{tabular} & \begin{tabular}[c]{@{}c@{}}0.294\\ 0.080\end{tabular} & \begin{tabular}[c]{@{}c@{}}0.577\\ 0.083\end{tabular} & \begin{tabular}[c]{@{}c@{}}0.960\\ 0.020\end{tabular} & \begin{tabular}[c]{@{}c@{}}0.315\\ 0.069\end{tabular} \\ \hline
\end{tabular}}
\end{table} 

From the results in Table \ref{comparetable}, the performance of the methods considered here improved with the increase in sample size. 
We observe that the PSRFR method can identify the whole subspace more accurately and estimate each direction well compared with the PHD and IHT methods, and the IHT is more accurate at recognizing the first direction.

In addition to the PHD and IHT methods, we further consider comparing the PSRFR method to the SIR, SAVE, and IRE methods under three different models with normally distributed predictor variables studied in Example 3 of \cite{zhu2006sliced}, \cite{Li1991}, and Example 3 of \cite{chen2015diagnostic}: 
\begin{itemize}  
\item {\bf Model [N3]}
\begin{equation*}
	Y=(4+\beta_{1}^\top\bm{X})\cdot (\beta_{2}^\top\bm{X}+2)+\sigma\varepsilon.
\end{equation*}
\item {\bf Model [N4]}
\begin{equation*}
Y=\beta_{1}^\top\bm{X}/\left\{0.5+(\beta_{2}^\top\bm{X}+3)^{2}\right\}+\sigma\varepsilon.
\end{equation*}
\item {\bf Model [N5]}
\begin{equation*}
	Y=\left( \beta_{1}^\top\bm{X}\right)^{2}+\left | \beta_{2}^\top\bm{X} \right | +\sigma\varepsilon.
\end{equation*}
\end{itemize}
We consider $\sigma=0.5$, the aforementioned settings for $\bm{X}$, $\sigma$, $\beta_1$, and $\beta_2$, and the number of slices to be $H = 10$. 
 The SIR, SAVE, and IRE methods are also implemented in the R package {\tt dr} \citep{drpackage2002}.  Table 2 reports the averages and standard deviations of the trace correlation $R$ based on 1000 simulations.
\begin{table}[!ht]
	\centering
	\caption{Averages and standard deviations of the trace correlation $R$ for the PSRFR, PHD, SIR, SAVE, IRE, and IHT methods for models {\bf [N3]}, {\bf [N4]}, and  {\bf [N5]} based on 1000 simulations with different sample sizes.}
 \label{TNormal}
        \setlength{\tabcolsep}{8pt} % Default value: 6pt
        \renewcommand{\arraystretch}{1.1}

\footnotesize{
\begin{tabular}{cccccccc}
\hline
\multicolumn{1}{c}{} & Method & PSRFR &   PHD & SIR & SAVE & IRE & IHT \\ \hline
\multicolumn{8}{c}{{\bf Model [N3]}} \\ \hline
\multirow{2}{*}{$n=100$} & Average & 0.895 & 0.946 & 0.802 & 0.347 & 0.625 & 0.970 \\
 & SD & 0.074 & 0.041 & 0.137 & 0.166 & 0.110 & 0.067 \\ 
\multirow{2}{*}{$n=300$} & Average & 0.968 & 0.986 & 0.941 & 0.682 & 0.811 & 0.983 \\
 & SD & 0.020 & 0.006 & 0.074 & 0.143 & 0.103 & 0.070 \\ 
\multirow{2}{*}{$n=500$} & Average & 0.980 & 0.992 & 0.976 & 0.712 & 0.889 & 0.983 \\
 & SD & 0.012 & 0.003 & 0.018 & 0.149 & 0.059 & 0.078 \\ \hline
\multicolumn{8}{c}{{\bf Model [N4]}} \\ \hline
\multirow{2}{*}{$n=100$} & Average & 0.857 & 0.695 & 0.625 & 0.421 & 0.313 & 0.581 \\
 & SD & 0.092 & 0.171 & 0.156 & 0.165 & 0.141 & 0.094 \\ 
\multirow{2}{*}{$n=300$} & Average & 0.943 & 0.900 & 0.800 & 0.519 & 0.537 & 0.585 \\
 & SD & 0.035 & 0.089 & 0.121 & 0.174 & 0.152 & 0.090 \\ 
\multirow{2}{*}{$n=500$} & Average & 0.961 & 0.950 & 0.867 & 0.606 & 0.667 & 0.586 \\
 & SD & 0.020 & 0.045 & 0.102 & 0.164 & 0.144 & 0.092 \\ \hline
\multicolumn{8}{c}{{\bf Model [N5]}} \\ \hline
\multirow{2}{*}{$n=100$} & Average & 0.943 & 0.901 & 0.431 & 0.759 & 0.172 & 0.542 \\
 & SD & 0.046 & 0.105 & 0.176 & 0.143 & 0.117 & 0.136 \\ 
\multirow{2}{*}{$n=300$} & Average & 0.980 & 0.982 & 0.426 & 0.978 & 0.173 & 0.550 \\
 & SD & 0.014 & 0.010 & 0.174 & 0.013 & 0.121 & 0.135 \\ 
\multirow{2}{*}{$n=500$} & Average & 0.988 & 0.990 & 0.431 & 0.990 & 0.177 & 0.543 \\
 & SD & 0.008 & 0.006 & 0.173 & 0.005 & 0.122 & 0.133 \\ \hline
\end{tabular}}
\end{table}

From Table \ref{TNormal}, once again, the performance of the methods considered here improved with the increase in sample size. The PSRFR method performs well in almost all the settings considered here, with different variances of each predictor component. Compared to the PHD and IHT methods, the PSRFR underperforms under model {\bf [N3]}, especially for sample size $n = 100$, since model {\bf [N3]} satisfies the assumptions for the PHD method is developed based on the normality of the predictor variables, and the IHT method is based on the PHD under the elliptically distributed predictor variables assumption.
The PSRFR method performs better than the other methods considered here under models {\bf [N4]} and {\bf [N5]}, which indicates that the PSRFR method is an effective method for estimating $\mathrm{Span}(\text{B})$ in the multivariate normal case with different variances of the predictor variables.

\subsubsection{Non-normal distributions}
\label{sec:312}

In this subsection, we consider the predictor variables following different non-normal multivariate elliptical distributions. Specifically, the multivariate Student's $t$ and multivariate power exponential distributions are considered: 
\begin{itemize}
\item[] {\bf Multivariate Student's $t$ distribution} \citep{kotz2004multivariate}: \\ 
A $p$-dimensional random vector $\bm{X}$ is said to be distributed as a multivariate Student's t distribution with degrees of freedom $\nu$, mean vector $\mu$, and positive-definite symmetric matrix $\Sigma$ if its joint probability density function is given by
\begin{eqnarray*}
    f_{t}(\bm{X}) & = & \frac{\Gamma((\nu+p)/2)}{(\pi\nu)^{p/2}\Gamma(\nu/2)|\Sigma|^{1/2}} \\
    & & \times \left[1+\frac{1}{\nu}(\bm{X}-\mu)^\top \Sigma^{-1}(\bm{X}-\mu)\right]^{-(\nu+p)/2},~~\bm{X} \in \mathbb{R}^{p}.
\end{eqnarray*}
As $\nu\rightarrow \infty$, the limiting form is the multivariate normal distribution. Hence, multivariate Student's $t$ distribution with small degrees of freedom deviates significantly from multivariate normal distribution, especially in the tail areas. In the special case of $\nu =1$, the multivariate Student's $t$ distribution is a multivariate Cauchy distribution. Notice that the variance-covariance matrix of the multivariate Student's $t$ distribution is given by $\nu/(\nu-2)\Sigma$ for $\nu>2$. Hence, for multivariate Student's $t$ distribution with degrees of freedom 1 and 2, the variance-covariance matrix does not exist. 
\item[] {\bf Multivariate power exponential distribution} \citep{gomez1998multivariate}:\\
A $p$-dimensional random vector $\bm{X}$ is said to be distributed as a multivariate power exponential distribution with kurtosis parameters $\beta>0$, mean vector $\mu$ and positive-definite symmetric matrix $\Sigma$ if its joint probability density function is given by
\begin{eqnarray*}
    f_{PE}(\bm{X})& = & \frac{p\Gamma(p/2)}{\Gamma(1+p/2\beta)\pi^{p/2}2^{1+p/2\beta}|\Sigma|^{1/2}}\\ & & 
    \times \textup{exp}\left[-\frac{1}{2}\left((\bm{X}-\mu)^\top \Sigma^{-1}(\bm{X}-\mu)\right)^\beta\right],~~\bm{X} \in \mathbb{R}^{p}.
\end{eqnarray*}
In particular, the multivariate Laplace and multivariate normal distributions are special cases of the multivariate power exponential distribution when the kurtosis parameter $\beta=0.5$ and $\beta=1$, respectively. Therefore, the kurtosis parameter $\beta$ can be viewed as the disparity between power exponential distribution and normal distribution.
\end{itemize} 

In the simulation study, we consider the predictor variables following the multivariate Student's $t$ distributions with degrees of freedom $\nu = 2$ and 3, the multivariate Cauchy distribution (i.e., multivariate Student's $t$ distributions with degree of freedom $\nu = 1$), and the multivariate power exponential distribution with kurtosis parameters $\beta = 0.5$ and 5. We utilize the R packages {\tt mvtnorm} \citep{mvtnorm2009} and {\tt LaplacesDemon} \citep{LaplacesDemon2021} to simulate random vectors from the multivariate Student's $t$ and the multivariate power exponential distributions, respectively.

%We consider three elliptical distributions and five specific settings, including the multivariate Student's $t$ distribution with degrees of freedom 2 and 3, multivariate Cauchy distribution, and multivariate power exponential distribution with kurtosis parameters 0.5 and 5. 
%{\clr In order to explain the meanings of the parameter, such as the kurtosis parameter,  we give the mathematical forms of above distributions. 
%\begin{definition}
%{\clr We use the “rmvt” function of “mvtnorm” package in R to generate random numbers of multivariate Student's t distribution, and use “rmvc” and “rmvpe” functions of “LaplacesDemon” package in R to generate random numbers of multivariate Cauchy distribution and multivariate power exponential distribution respectively.}
%}

The following four models under non-normal elliptical distributed predictor variables are considered: 
\begin{itemize} 
\item {\bf Model [NN1]}: 
\begin{equation*}
Y=\left(4+\beta_{1}^\top\bm{X}\right)+ \left(\beta_{2}^\top\bm{X}+2\right)\cdot \sigma\varepsilon^2.
\end{equation*}
This model is from Example 2 of \cite{zhu2006sliced}.
\item {\bf Model [NN2]}: 
\begin{equation*}
Y=\left(\left|4+\beta_{1}^\top\bm{X}\right|\right)^{1/2}\cdot \left(\left|\beta_{2}^\top\bm{X}+2\right|\right)^{1/2} +\sigma\varepsilon. 
\end{equation*}
This model is based on model {\bf [NN1]} with a slow-growing power function of degree $1/2$. 
\item {\bf Model [NN3]}: 
\begin{equation*}
	Y=\left(\left|\beta_{1}^\top\bm{X}\right|\right)^{1/2} +\left(\left|\beta_{2}^\top\bm{X}\cdot \varepsilon\right|\right)^{1/2} +\sigma\varepsilon. 
\end{equation*}
This model is also based on model {\bf [NN1]} with a slow-growing power function of degree $1/2$. 
\item {\bf Model [NN4]}: 
\begin{equation*}
Y=0.4\cdot \left(\beta_{1}^\top\bm{X}\right)+3\cdot \mathrm{sin}\left ( \beta_{2}^\top\bm{X}/4 \right )+\sigma\varepsilon.
\end{equation*}
This model is motivated by Example 1 of \cite{li2007directional}.
\end{itemize} 
%Model (B1) comes from Example 2 of \cite{zhu2006sliced}. In order to avoid calculating out of bounds because of the heavy-tailed distribution, especially multivariate Cauchy distribution, Model (B2) and Model (B3) are motivated by Model (B1), where we adopt the slow-growing function power function of degree 0.5. Model (B4) changes the quadratic term into a first term motivated by Example 1 of \cite{li2007directional}. 

Following the settings in Section \ref{sec:311}, we consider $\mathrm{dim}(\bm{X}) = 10$, $\sigma=0.5$, $\varepsilon\sim \mathrm{N}(0,1)$, $\beta_{1}=(1,0,\dots,0)$ and $\beta_{2}=(0,1,\dots,0)$. We set $\mu = 0$ and $\Sigma = \Sigma_{ellp}$  is a diagonal matrix with elements $(1, 6, 11, 16, 21, 26, 31, 36, 41, 46)$ in the multivariate Student's $t$ and power exponential distributions. 
%For the other three settings, $\mu = 0$ and $\Sigma_{ellp}$ is a diagonal matrix whose elements increase sequentially from 1 with the step size 5.  
The averages and standard deviations of the trace correlation $R$ for different methods based on 1000 simulations are reported in Tables \ref{TB1}--\ref{TB4} for models {\bf [NN1]} -- {\bf [NN4]}, respectively.
%report the mean and standard deviation of $R$ after 1000 replicates.

The results in Tables \ref{TB1}--\ref{TB4} show that the PSRFR method outperforms other methods in most of the models and settings when the predictor variables follow a non-normal elliptical distribution, especially when the distribution has heavier tails compared to the multivariate normal distribution (i.e., the multivariate Student's $t$  distribution with small degree of freedom $\nu$, and the multivariate power exponential distribution with large kurtosis parameter $\beta$). 
%These distributions have heavy tails compared to the normal distribution, particularly the Student's $t$  distribution with small degrees of freedom, the power exponential distribution with a large kurtosis parameter, and the Cauchy distribution. 
The PHD and SAVE methods exhibit comparable performance to the PSRFR method when the predictor variables follow the multivariate power exponential distribution with kurtosis parameter $\beta = 0.5$ since the multivariate power exponential distribution with small kurtosis parameter behaves similarly to multivariate normal distribution. 
%erform worse relative to others because they are obtained under the normal assumption. We can also find that other methods perform slightly better with the power exponential distribution than with the others. The reason is that the tail of the power exponential distribution when the kurtosis is 0.5 is very close to that of the normal distribution.

\begin{table}[!ht]
\centering
\caption{Averages and standard deviations of the trace correlation $R$ for the PSRFR, PHD, SIR, SAVE, IRE, and IHT methods for model {\bf [NN1]} based on 1000 simulations with different sample sizes.}
 \label{TB1}
 \setlength{\tabcolsep}{8.5pt} % Default value: 6pt
\renewcommand{\arraystretch}{1.0}
{\footnotesize
%\hspace*{8.5em}
\begin{tabular}{cccccccclllll}
\cline{1-8}
\multicolumn{1}{c}{{\bf Model [NN1]}} & Method: & PSRFR & PHD & SIR & SAVE & IRE & IHT & & & & & \\ \cline{1-8}
\multicolumn{8}{c}{Multivariate Student's $t$ with $\nu = 3$} & & & & & \\ \cline{1-8}
\multirow{2}{*}{$n=100$} & Average & 0.858 & 0.671 & 0.801 & 0.571 & 0.494 & 0.795 & & & & & \\
 & SD & 0.113 & 0.158 & 0.139 & 0.191 & 0.095 & 0.113 & & & & & \\ 
\multirow{2}{*}{$n=300$} & Average & 0.902 & 0.686 & 0.856 & 0.605 & 0.600 & 0.875 & & & & & \\
 & SD & 0.084 & 0.155 & 0.133 & 0.194 & 0.116 & 0.095 & & & & & \\ 
\multirow{2}{*}{$n=500$} & Average & 0.907 & 0.683 & 0.900 & 0.631 & 0.658 & 0.907 & & & & & \\
 & SD & 0.087 & 0.160 & 0.112 & 0.187 & 0.125 & 0.078 & & & & & \\ \cline{1-8}
\multicolumn{8}{c}{Multivariate Student's $t$ with $\nu = 2$} & & & & & \\ \cline{1-8}
\multirow{2}{*}{$n=100$} & Average & 0.826 & 0.669 & 0.802 & 0.582 & 0.439 & 0.725 & & & & & \\
 & SD & 0.127 & 0.161 & 0.140 & 0.188 & 0.108 & 0.136 & & & & & \\ 
\multirow{2}{*}{$n=300$} & Average & 0.835 & 0.678 & 0.820 & 0.575 & 0.509 & 0.784 & & & & & \\
 & SD & 0.127 & 0.165 & 0.139 & 0.190 & 0.118 & 0.128 & & & & & \\ 
\multirow{2}{*}{$n=500$} & Average & 0.834 & 0.678 & 0.827 & 0.575 & 0.531 & 0.798 & & & & & \\
 & SD & 0.126 & 0.165 & 0.144 & 0.195 & 0.126 & 0.128 & & & & & \\ \cline{1-8}
\multicolumn{8}{c}{Multivariate Cauchy} & & & & & \\ \cline{1-8}
\multirow{2}{*}{$n=100$} & Average & 0.755 & 0.650 & 0.785 & 0.648 & 0.278 & 0.571 & & & & & \\
 & SD & 0.141 & 0.181 & 0.155 & 0.180 & 0.117 & 0.149 & & & & & \\ 
\multirow{2}{*}{$n=300$} & Average & 0.750 & 0.647 & 0.758 & 0.634 & 0.307 & 0.576 & & & & & \\
 & SD & 0.146 & 0.174 & 0.137 & 0.187 & 0.118 & 0.149 & & & & & \\ 
\multirow{2}{*}{$n=500$} & Average & 0.746 & 0.636 & 0.751 & 0.621 & 0.314 & 0.581 & & & & & \\
 & SD & 0.144 & 0.177 & 0.134 & 0.191 & 0.121 & 0.139 & & & & & \\ \cline{1-8}
\multicolumn{8}{c}{Multivariate power exponential with $\beta = 0.5$} & \multicolumn{1}{c}{} & \multicolumn{1}{c}{} & \multicolumn{1}{c}{} & \multicolumn{1}{c}{} & \multicolumn{1}{c}{} \\ \cline{1-8}
\multirow{2}{*}{$n=100$} & Average & 0.901 & 0.687 & 0.803 & 0.570 & 0.547 & 0.877 & & & & & \\
 & SD & 0.086 & 0.155 & 0.140 & 0.198 & 0.094 & 0.074 & \multicolumn{1}{c}{} & & & & \\ 
\multirow{2}{*}{$n=300$} & Average & 0.958 & 0.703 & 0.917 & 0.832 & 0.705 & 0.944 & & & & & \\
 & SD & 0.034 & 0.149 & 0.101 & 0.125 & 0.122 & 0.070 & \multicolumn{1}{c}{} & & & & \\ 
\multirow{2}{*}{$n=500$} & Average & 0.969 & 0.707 & 0.954 & 0.815 & 0.787 & 0.958 & & & & & \\
 & SD & 0.027 & 0.150 & 0.074 & 0.136 & 0.114 & 0.080 & \multicolumn{1}{c}{} & & & & \\ \cline{1-8}
\multicolumn{8}{c}{Multivariate power exponential with $\beta = 5$} & \multicolumn{1}{c}{} & \multicolumn{1}{c}{} & \multicolumn{1}{c}{} & \multicolumn{1}{c}{} & \multicolumn{1}{c}{} \\ \cline{1-8}
\multirow{2}{*}{$n=100$} & Average & 0.947 & 0.627 & 0.805 & 0.585 & 0.454 & 0.683 & & & & & \\
 & SD & 0.054 & 0.179 & 0.127 & 0.188 & 0.120 & 0.134 & \multicolumn{1}{c}{} & & & & \\ 
\multirow{2}{*}{$n=300$} & Average & 0.983 & 0.629 & 0.857 & 0.799 & 0.665 & 0.819 & & & & & \\
 & SD & 0.015 & 0.174 & 0.115 & 0.140 & 0.121 & 0.099 & \multicolumn{1}{c}{} & & & & \\ 
\multirow{2}{*}{$n=500$} & Average & 0.990 & 0.634 & 0.918 & 0.797 & 0.772 & 0.875 & & & & & \\
 & SD & 0.007 & 0.181 & 0.086 & 0.129 & 0.108 & 0.076 & \multicolumn{1}{c}{} & & & & \\ \cline{1-8}
\end{tabular}}
\end{table}

\begin{table}[!ht]
	\centering
	\caption{Averages and standard deviations of the trace correlation $R$ for the PSRFR, PHD, SIR, SAVE, IRE, and IHT methods for model {\bf [NN2]} based on 1000 simulations with different sample sizes.}
 \label{TB2}
        \setlength{\tabcolsep}{8.5pt} % Default value: 6pt
        \renewcommand{\arraystretch}{1.0}
{\footnotesize
%\hspace*{8.5em}
\begin{tabular}{ccccccccccccc}
\cline{1-8}
\multicolumn{1}{c}{{\bf Model [NN2]}} & Method: & PSRFR & PHD & SIR & SAVE & IRE & IHT & & & & & \\ \cline{1-8}
\multicolumn{8}{c}{Multivariate Student's $t$ with $\nu = 3$} & & & & & \\ \cline{1-8}
\multirow{2}{*}{$n=100$} & Average & 0.955 & 0.770 & 0.773 & 0.652 & 0.378 & 0.723 & & & & & \\
 & SD & 0.037 & 0.140 & 0.130 & 0.177 & 0.119 & 0.149 & & & & & \\ 
\multirow{2}{*}{$n=300$} & Average & 0.964 & 0.812 & 0.757 & 0.700 & 0.474 & 0.776 & & & & & \\
 & SD & 0.042 & 0.123 & 0.139 & 0.155 & 0.094 & 0.151 & & & & & \\ 
\multirow{2}{*}{$n=500$} & Average & 0.963 & 0.825 & 0.779 & 0.727 & 0.515 & 0.814 & & & & & \\
 & SD & 0.037 & 0.113 & 0.139 & 0.145 & 0.094 & 0.136 & & & & & \\ \cline{1-8}
\multicolumn{8}{c}{Multivariate Student's $t$ with $\nu = 2$} & & & & & \\ \cline{1-8}
\multirow{2}{*}{$n=100$} & Average & 0.930 & 0.767 & 0.760 & 0.687 & 0.306 & 0.606 & & & & & \\
 & SD & 0.062 & 0.135 & 0.139 & 0.173 & 0.131 & 0.150 & & & & & \\ 
\multirow{2}{*}{$n=300$} & Average & 0.922 & 0.806 & 0.802 & 0.731 & 0.371 & 0.605 & & & & & \\
 & SD & 0.077 & 0.124 & 0.119 & 0.158 & 0.117 & 0.152 & & & & & \\ 
\multirow{2}{*}{$n=500$} & Average & 0.909 & 0.813 & 0.814 & 0.741 & 0.406 & 0.606 & & & & & \\
 & SD & 0.089 & 0.121 & 0.112 & 0.153 & 0.118 & 0.153 & & & & & \\ \cline{1-8}
\multicolumn{8}{c}{Multivariate Cauchy} & & & & & \\ \cline{1-8}
\multirow{2}{*}{$n=100$} & Average & 0.832 & 0.708 & 0.662 & 0.695 & 0.142 & 0.460 & & & & & \\
 & SD & 0.125 & 0.177 & 0.188 & 0.178 & 0.093 & 0.171 & & & & & \\ 
\multirow{2}{*}{$n=300$} & Average & 0.829 & 0.723 & 0.664 & 0.732 & 0.170 & 0.486 & & & & & \\
 & SD & 0.125 & 0.164 & 0.180 & 0.157 & 0.102 & 0.150 & & & & & \\ 
\multirow{2}{*}{$n=500$} & Average & 0.818 & 0.725 & 0.698 & 0.751 & 0.193 & 0.479 & & & & & \\
 & SD & 0.134 & 0.170 & 0.174 & 0.159 & 0.103 & 0.160 & & & & & \\ \cline{1-8}
\multicolumn{8}{c}{Multivariate power exponential with $\beta = 0.5$} & \multicolumn{1}{c}{} & \multicolumn{1}{c}{} & \multicolumn{1}{c}{} & \multicolumn{1}{c}{} & \multicolumn{1}{c}{} \\ \cline{1-8}
\multirow{2}{*}{$n=100$} & Average & 0.977 & 0.782 & 0.774 & 0.644 & 0.445 & 0.831 & & & & & \\
 & SD & 0.015 & 0.142 & 0.133 & 0.172 & 0.100 & 0.100 & \multicolumn{1}{c}{} & & & & \\ 
\multirow{2}{*}{$n=300$} & Average & 0.993 & 0.772 & 0.748 & 0.772 & 0.531 & 0.932 & & & & & \\
 & SD & 0.004 & 0.152 & 0.143 & 0.140 & 0.075 & 0.048 & \multicolumn{1}{c}{} & & & & \\ 
\multirow{2}{*}{$n=500$} & Average & 0.996 & 0.745 & 0.745 & 0.830 & 0.548 & 0.954 & & & & & \\
 & SD & 0.002 & 0.153 & 0.152 & 0.147 & 0.078 & 0.048 & \multicolumn{1}{c}{} & & & & \\ \cline{1-8}
\multicolumn{8}{c}{Multivariate power exponential with $\beta = 5$} & \multicolumn{1}{c}{} & \multicolumn{1}{c}{} & \multicolumn{1}{c}{} & \multicolumn{1}{c}{} & \multicolumn{1}{c}{} \\ \cline{1-8}
\multirow{2}{*}{$n=100$} & Average & 0.961 & 0.644 & 0.765 & 0.580 & 0.487 & 0.936 & & & & & \\
 & SD & 0.034 & 0.183 & 0.144 & 0.192 & 0.076 & 0.080 & \multicolumn{1}{c}{} & & & & \\ 
\multirow{2}{*}{$n=300$} & Average & 0.988 & 0.687 & 0.780 & 0.766 & 0.545 & 0.979 & & & & & \\
 & SD & 0.009 & 0.170 & 0.151 & 0.132 & 0.068 & 0.039 & \multicolumn{1}{c}{} & & & & \\ 
\multirow{2}{*}{$n=500$} & Average & 0.993 & 0.708 & 0.708 & 0.764 & 0.553 & 0.985 & & & & & \\
 & SD & 0.005 & 0.161 & 0.161 & 0.145 & 0.072 & 0.045 & \multicolumn{1}{c}{} & & & & \\ \cline{1-8}
\end{tabular}}
\end{table}

\begin{table}[!ht]
	\centering
	\caption{Averages and standard deviations of the trace correlation $R$ for the PSRFR, PHD, SIR, SAVE, IRE, and IHT methods for model {\bf [NN3]} based on 1000 simulations with different sample sizes.}
 \label{TB3}
        \setlength{\tabcolsep}{8.5pt} % Default value: 6pt
\renewcommand{\arraystretch}{1.0}
{\footnotesize
%\hspace*{8.5em}
\begin{tabular}{cccccccclllll}
\cline{1-8}
\multicolumn{1}{c}{{\bf Model [NN3]}} & Method: & PSRFR & PHD & SIR & SAVE & IRE & IHT & & & & & \\ \cline{1-8}
\multicolumn{8}{c}{Multivariate Student's $t$ with $\nu = 3$} & & & & & \\ \cline{1-8}
\multirow{2}{*}{$n=100$} & Average & 0.956 & 0.828 & 0.633 & 0.709 & 0.186 & 0.527 & & & & & \\
 & SD & 0.046 & 0.125 & 0.180 & 0.171 & 0.113 & 0.158 & & & & & \\ 
\multirow{2}{*}{$n=300$} & Average & 0.978 & 0.884 & 0.625 & 0.790 & 0.165 & 0.533 & & & & & \\
 & SD & 0.024 & 0.107 & 0.177 & 0.146 & 0.104 & 0.148 & & & & & \\ 
\multirow{2}{*}{$n=500$} & Average & 0.982 & 0.897 & 0.615 & 0.822 & 0.170 & 0.536 & & & & & \\
 & SD & 0.018 & 0.096 & 0.187 & 0.136 & 0.104 & 0.154 & & & & & \\ \cline{1-8}
\multicolumn{8}{c}{Multivariate Student's $t$ with $\nu = 2$} & & & & & \\ \cline{1-8}
\multirow{2}{*}{$n=100$} & Average & 0.937 & 0.797 & 0.632 & 0.710 & 0.169 & 0.509 & & & & & \\
 & SD & 0.070 & 0.136 & 0.177 & 0.165 & 0.105 & 0.161 & & & & & \\ 
\multirow{2}{*}{$n=300$} & Average & 0.962 & 0.827 & 0.622 & 0.754 & 0.158 & 0.516 & & & & & \\
 & SD & 0.041 & 0.128 & 0.182 & 0.157 & 0.103 & 0.158 & & & & & \\ 
\multirow{2}{*}{$n=500$} & Average & 0.965 & 0.833 & 0.631 & 0.772 & 0.158 & 0.531 & & & & & \\
 & SD & 0.041 & 0.127 & 0.181 & 0.155 & 0.099 & 0.140 & & & & & \\ \cline{1-8}
\multicolumn{8}{c}{Multivariate Cauchy} & & & & & \\ \cline{1-8}
\multirow{2}{*}{$n=100$} & Average & 0.854 & 0.717 & 0.624 & 0.692 & 0.131 & 0.469 & & & & & \\
 & SD & 0.118 & 0.170 & 0.194 & 0.173 & 0.086 & 0.171 & & & & & \\ 
\multirow{2}{*}{$n=300$} & Average & 0.869 & 0.712 & 0.600 & 0.724 & 0.120 & 0.491 & & & & & \\
 & SD & 0.118 & 0.165 & 0.183 & 0.167 & 0.081 & 0.156 & & & & & \\ 
\multirow{2}{*}{$n=500$} & Average & 0.872 & 0.725 & 0.612 & 0.743 & 0.119 & 0.476 & & & & & \\
 & SD & 0.114 & 0.169 & 0.183 & 0.162 & 0.018 & 0.170 & & & & & \\ \cline{1-8}
\multicolumn{8}{c}{Multivariate power exponential with $\beta = 0.5$} & \multicolumn{1}{c}{} & \multicolumn{1}{c}{} & \multicolumn{1}{c}{} & \multicolumn{1}{c}{} & \multicolumn{1}{c}{} \\ \cline{1-8}
\multirow{2}{*}{$n=100$} & Average & 0.975 & 0.883 & 0.637 & 0.754 & 0.195 & 0.541 & & & & & \\
 & SD & 0.017 & 0.116 & 0.180 & 0.152 & 0.118 & 0.147 & \multicolumn{1}{c}{} & & & & \\ 
\multirow{2}{*}{$n=300$} & Average & 0.992 & 0.979 & 0.623 & 0.948 & 0.185 & 0.535 & & & & & \\
 & SD & 0.005 & 0.029 & 0.182 & 0.070 & 0.114 & 0.155 & \multicolumn{1}{c}{} & & & & \\ 
\multirow{2}{*}{$n=500$} & Average & 0.995 & 0.990 & 0.622 & 0.986 & 0.183 & 0.534 & & & & & \\
 & SD & 0.003 & 0.007 & 0.184 & 0.018 & 0.115 & 0.163 & \multicolumn{1}{c}{} & & & & \\ \cline{1-8}
\multicolumn{8}{c}{Multivariate power exponential with $\beta = 5$} & \multicolumn{1}{c}{} & \multicolumn{1}{c}{} & \multicolumn{1}{c}{} & \multicolumn{1}{c}{} & \multicolumn{1}{c}{} \\ \cline{1-8}
\multirow{2}{*}{$n=100$} & Average & 0.967 & 0.679 & 0.624 & 0.713 & 0.192 & 0.518 & & & & & \\
 & SD & 0.023 & 0.172 & 0.176 & 0.165 & 0.112 & 0.144 & \multicolumn{1}{c}{} & & & & \\ 
\multirow{2}{*}{$n=300$} & Average & 0.990 & 0.812 & 0.627 & 0.879 & 0.197 & 0.517 & & & & & \\
 & SD & 0.006 & 0.200 & 0.184 & 0.119 & 0.115 & 0.150 & \multicolumn{1}{c}{} & & & & \\ 
\multirow{2}{*}{$n=500$} & Average & 0.994 & 0.917 & 0.624 & 0.964 & 0.196 & 0.519 & & & & & \\
 & SD & 0.004 & 0.159 & 0.175 & 0.062 & 0.110 & 0.155 & \multicolumn{1}{c}{} & & & & \\ \cline{1-8}
\end{tabular}}
\end{table}

\begin{table}[!ht]
	\centering
	\caption{Averages and standard deviations of the trace correlation $R$ for the PSRFR, PHD, SIR, SAVE, IRE, and IHT methods for model {\bf [NN4]} based on 1000 simulations with different sample sizes.}
 \label{TB4}
        \setlength{\tabcolsep}{8.5pt} % Default value: 6pt
\renewcommand{\arraystretch}{1.0}
\footnotesize{
%\hspace*{8.5em}
\begin{tabular}{cccccccclllll}
\cline{1-8}
\multicolumn{1}{c}{{\bf Model [NN4]}} & Method: & PSRFR & PHD & SIR & SAVE & IRE & IHT & & & & & \\ \cline{1-8}
\multicolumn{8}{c}{Multivariate Student's $t$ with $\nu = 3$} & & & & & \\ \cline{1-8}
\multirow{2}{*}{$n=100$} & Average & 0.915 & 0.708 & 0.803 & 0.570 & 0.485 & 0.901 & & & & & \\
 & SD & 0.084 & 0.144 & 0.141 & 0.190 & 0.112 & 0.084 & & & & & \\ 
\multirow{2}{*}{$n=300$} & Average & 0.942 & 0.728 & 0.881 & 0.583 & 0.609 & 0.946 & & & & & \\
 & SD & 0.062 & 0.139 & 0.129 & 0.178 & 0.120 & 0.062 & & & & & \\ 
\multirow{2}{*}{$n=500$} & Average & 0.950 & 0.724 & 0.937 & 0.583 & 0.701 & 0.957 & & & & & \\
 & SD & 0.061 & 0.140 & 0.094 & 0.182 & 0.117 & 0.070 & & & & & \\ \cline{1-8}
\multicolumn{8}{c}{Multivariate Student's $t$ with $\nu = 2$} & & & & & \\ \cline{1-8}
\multirow{2}{*}{$n=100$} & Average & 0.866 & 0.703 & 0.798 & 0.570 & 0.422 & 0.793 & & & & & \\
 & SD & 0.114 & 0.151 & 0.139 & 0.181 & 0.122 & 0.128 & & & & & \\ 
\multirow{2}{*}{$n=300$} & Average & 0.862 & 0.713 & 0.872 & 0.539 & 0.557 & 0.815 & & & & & \\
 & SD & 0.120 & 0.150 & 0.123 & 0.173 & 0.139 & 0.128 & & & & & \\ 
\multirow{2}{*}{$n=500$} & Average & 0.850 & 0.699 & 0.914 & 0.548 & 0.636 & 0.817 & & & & & \\
 & SD & 0.134 & 0.148 & 0.105 & 0.187 & 0.143 & 0.138 & & & & & \\ \cline{1-8}
\multicolumn{8}{c}{Multivariate Cauchy} & & & & & \\ \cline{1-8}
\multirow{2}{*}{$n=100$} & Average & 0.752 & 0.657 & 0.783 & 0.626 & 0.271 & 0.534 & & & & & \\
 & SD & 0.138 & 0.174 & 0.127 & 0.183 & 0.108 & 0.089 & & & & & \\ 
\multirow{2}{*}{$n=300$} & Average & 0.730 & 0.648 & 0.770 & 0.617 & 0.337 & 0.521 & & & & & \\
 & SD & 0.143 & 0.174 & 0.125 & 0.187 & 0.135 & 0.082 & & & & & \\ 
\multirow{2}{*}{$n=500$} & Average & 0.727 & 0.661 & 0.754 & 0.599 & 0.353 & 0.520 & & & & & \\
 & SD & 0.143 & 0.169 & 0.127 & 0.191 & 0.139 & 0.080 & & & & & \\ \cline{1-8}
\multicolumn{8}{c}{Multivariate power exponential with $\beta = 0.5$} & \multicolumn{1}{c}{} & \multicolumn{1}{c}{} & \multicolumn{1}{c}{} & \multicolumn{1}{c}{} & \multicolumn{1}{c}{} \\ \cline{1-8}
\multirow{2}{*}{$n=100$} & Average & 0.953 & 0.719 & 0.840 & 0.559 & 0.576 & 0.955 & & & & & \\
 & SD & 0.049 & 0.142 & 0.146 & 0.198 & 0.104 & 0.060 & \multicolumn{1}{c}{} & & & & \\ 
\multirow{2}{*}{$n=300$} & Average & 0.985 & 0.750 & 0.945 & 0.784 & 0.729 & 0.982 & & & & & \\
 & SD & 0.013 & 0.126 & 0.090 & 0.138 & 0.114 & 0.048 & \multicolumn{1}{c}{} & & & & \\ 
\multirow{2}{*}{$n=500$} & Average & 0.991 & 0.753 & 0.978 & 0.773 & 0.819 & 0.987 & & & & & \\
 & SD & 0.006 & 0.128 & 0.054 & 0.160 & 0.095 & 0.050 & \multicolumn{1}{c}{} & & & & \\ \cline{1-8}
\multicolumn{8}{c}{Multivariate power exponential with $\beta = 5$} & \multicolumn{1}{c}{} & \multicolumn{1}{c}{} & \multicolumn{1}{c}{} & \multicolumn{1}{c}{} & \multicolumn{1}{c}{} \\ \cline{1-8}
\multirow{2}{*}{$n=100$} & Average & 0.972 & 0.626 & 0.767 & 0.576 & 0.493 & 0.942 & & & & & \\
 & SD & 0.019 & 0.185 & 0.149 & 0.198 & 0.075 & 0.060 & \multicolumn{1}{c}{} & & & & \\ 
\multirow{2}{*}{$n=300$} & Average & 0.992 & 0.619 & 0.757 & 0.773 & 0.540 & 0.978 & & & & & \\
 & SD & 0.005 & 0.178 & 0.155 & 0.130 & 0.064 & 0.046 & \multicolumn{1}{c}{} & & & & \\ 
\multirow{2}{*}{$n=500$} & Average & 0.995 & 0.621 & 0.763 & 0.770 & 0.545 & 0.984 & & & & & \\
 & SD & 0.003 & 0.188 & 0.152 & 0.147 & 0.063 & 0.050 & \multicolumn{1}{c}{} & & & & \\ \cline{1-8}
\end{tabular} }
\end{table}

\subsection{Non-elliptical Distribution for Robust Analysis}
\label{sec3.2}

In this subsection, we perform a simulation study to investigate whether the PSRFR method can effectively identify $\mathrm{Span}(\text{B})$ when the predictor variables do not follow an elliptical distribution. Under the non-elliptical distribution situations, we consider comparing the proposed PSRFR method to the MAVE and the OPG methods proposed by \citet{xia2002adaptive} for identifying the central mean subspace. The MAVE and the OPG methods require a differentiable link function but have no strict distributional assumptions for the predictor variables. The OPG and MAVE methods are implemented in the R package {\tt MAVE} \citep{MAVE2021}. 

Before considering the non-elliptical distribution situations, we compare the 
OPG and MAVE methods to the PSRFR methods under model {\bf [NN1]} and {\bf [NN4]} when the predictor variables follow an elliptical distribution. Specifically, in Table \ref{TNot3MAOP}, we present the averages and standard deviations of the $\mathrm{trace}$ correlations based on 1000 simulations for PSRFR, OPG, and MAVE methods when the predictor variables follow a multivariate normal distribution or multivariate Student's $t$ distribution with degrees of freedom $\nu = 3$ with the setting described in Sections \ref{sec:311} and \ref{sec:312}. 
The results in Table \ref{TNot3MAOP} show that the PSRFR method performs better than the OPG and MAVE methods when the predictor variables follow an elliptical distribution.

For the non-elliptical distribution situation, the predictor variables $\bm{X}$ are generated from a mixture of multivariate normal and multivariate uniform distribution in $(-3, 3)$ with mixture probabilities 0.8 and 0.2, respectively, i.e.,
\begin{eqnarray}
\bm{X} \sim 0.8\mathrm{N}_{10}(0, \Sigma_{norm})+0.2 \mathrm{U}_{10}(-3,3),   
\label{mixnu}
\end{eqnarray}
where $\mathrm{U}_{10}(-3,3)$ denotes a 10-dimensional multivariate uniform distribution in $(-3, 3)$, each component of which independently follows uniform distribution from $-3$ to $3$. The following three models are considered:
\begin{itemize} 
\item {\bf Model [NE1]:}
\begin{equation*}
Y=\beta_{1}^\top\bm{X}/\left\{0.5+(\beta_{2}^\top\bm{X}+1.5)^{2}\right\}+\sigma\varepsilon. 
\end{equation*}
\item {\bf Model [NE2]:}
\begin{equation*}
Y=\beta_{1}^\top\bm{X}\cdot \left(\beta_{2}^\top\bm{X}+1\right) + \sigma\varepsilon. 
\end{equation*}
\item {\bf Model [NE3]:}
\begin{equation*}
Y=0.4\cdot \left(\beta_{1}^\top\bm{X}\right)+3\cdot \mathrm{sin}\left ( \beta_{1}^\top\bm{X}\bm{X}^\top\beta_{2}/4 \right )+\sigma\varepsilon.
\end{equation*}
\end{itemize} 
Following the settings in Sections \ref{sec:311} and \ref{sec:312}, we consider $\sigma=0.5$, $\varepsilon\sim \mathrm{N}(0,1)$, $\beta_{1}=(1,0,\dots,0)$,  $\beta_{2}=(0,1,\dots,0)$ and 
$\Sigma_{norm}$ is a diagonal matrix with elements $(1, \ldots, 10)$.

The averages and standard deviations of the trace correlations based on 1000 simulations for PSRFR, OPG, and MAVE methods when the predictor variables follow the distribution in Eq. \eqref{mixnu} are presented in Table \ref{TNonE}. From Table \ref{TNonE}, the PSRFR performs about 10\% worse relative to the OPG and MAVE method in terms of the trace correlation when $n = 100$, and the differences decrease 
to less than 5\% when the sample size increases to $n = 500$. %There is less than a 2 percent difference in performance for the PSRFR in 500 samples under models {\bf [NE2]} and {\bf [NE3]}. 
Although the OPG and MAVE methods perform better than the PSRFR method under the non-elliptical distribution situations, these simulation results demonstrate that the proposed PSRFR method is robust to the underlying distribution of the predictor variables. 
The proposed PSRFR method 
can effectively identify central subspaces, even the distribution of the predictor variables deviations from the elliptical distribution.

\begin{table}[!p]
\centering
\caption{
Averages and standard deviations of the trace correlation $R$ for the PSRFR, OPG, and MAVE methods for models {\bf [NN1]} and {\bf [NN4]} based on 1000 simulations with different sample sizes.}
\label{TNot3MAOP}
\setlength{\tabcolsep}{8pt} % Default value: 6pt
\renewcommand{\arraystretch}{1.1}
{\footnotesize
\begin{tabular}{cccccccc}
\hline
 & & \multicolumn{6}{c}{Multivariate Normal} \\
 &   & \multicolumn{3}{c}{\bf Model [NN1]} & \multicolumn{3}{c}{{\bf Model [NN4]}}  \\ 
\cline{2-8} & $n$ & 100 & 300& 500 & 100  & 300 & 500 \\ \hline
PSRFR                     & \begin{tabular}[c]{@{}c@{}}Average\\ SD\end{tabular} & \begin{tabular}[c]{@{}c@{}}0.840\\ 0.117\end{tabular} & \begin{tabular}[c]{@{}c@{}}0.945\\ 0.035\end{tabular} & \begin{tabular}[c]{@{}c@{}}0.966\\ 0.025\end{tabular} & \begin{tabular}[c]{@{}c@{}}0.921\\ 0.078\end{tabular} & \begin{tabular}[c]{@{}c@{}}0.976\\ 0.013\end{tabular} & \begin{tabular}[c]{@{}c@{}}0.987\\ 0.078\end{tabular} \\ 
OPG                     & \begin{tabular}[c]{@{}c@{}}Average\\ SD\end{tabular} & \begin{tabular}[c]{@{}c@{}}0.707\\ 0.123\end{tabular} & \begin{tabular}[c]{@{}c@{}}0.728\\ 0.128\end{tabular} & \begin{tabular}[c]{@{}c@{}}0.724\\ 0.130\end{tabular} & \begin{tabular}[c]{@{}c@{}}0.623\\ 0.103\end{tabular} & \begin{tabular}[c]{@{}c@{}}0.645\\ 0.127\end{tabular} & \begin{tabular}[c]{@{}c@{}}0.648\\ 0.122\end{tabular} \\ 
MAVE                    & \begin{tabular}[c]{@{}c@{}}Average\\ SD\end{tabular} & \begin{tabular}[c]{@{}c@{}}0.719\\ 0.131\end{tabular} & \begin{tabular}[c]{@{}c@{}}0.718\\ 0.128\end{tabular} & \begin{tabular}[c]{@{}c@{}}0.721\\ 0.146\end{tabular} & \begin{tabular}[c]{@{}c@{}}0.649\\ 0.125\end{tabular} & \begin{tabular}[c]{@{}c@{}}0.655\\ 0.121\end{tabular} & \begin{tabular}[c]{@{}c@{}}0.650\\ 0.127\end{tabular} \\ \hline

 & & \multicolumn{6}{c}{Multivariate Student's $t$ with $\nu = 3$} \\
 &   & \multicolumn{3}{c}{\bf Model [NN1]} & \multicolumn{3}{c}{{\bf Model [NN4]}}  \\ 
\cline{2-8} & $n$ & 100 & 300& 500 & 100  & 300 & 500 \\ \hline
PSRFR                     & \begin{tabular}[c]{@{}c@{}}Average\\ SD\end{tabular} & \begin{tabular}[c]{@{}c@{}}0.858\\ 0.121\end{tabular} & \begin{tabular}[c]{@{}c@{}}0.914\\ 0.072\end{tabular} & \begin{tabular}[c]{@{}c@{}}0.905\\ 0.084\end{tabular} & \begin{tabular}[c]{@{}c@{}}0.908\\ 0.091\end{tabular} & \begin{tabular}[c]{@{}c@{}}0.935\\ 0.075\end{tabular} & \begin{tabular}[c]{@{}c@{}}0.963\\ 0.030\end{tabular} \\ 
OPG                     & \begin{tabular}[c]{@{}c@{}}Average\\ SD\end{tabular} & \begin{tabular}[c]{@{}c@{}}0.682\\ 0.187\end{tabular} & \begin{tabular}[c]{@{}c@{}}0.751\\ 0.149\end{tabular} & \begin{tabular}[c]{@{}c@{}}0.763\\ 0.148\end{tabular} & \begin{tabular}[c]{@{}c@{}}0.748\\ 0.130\end{tabular} & \begin{tabular}[c]{@{}c@{}}0.757\\ 0.116\end{tabular} & \begin{tabular}[c]{@{}c@{}}0.748\\ 0.131\end{tabular} \\ 
MAVE                    & \begin{tabular}[c]{@{}c@{}}Average\\ SD\end{tabular} & \begin{tabular}[c]{@{}c@{}}0.843\\ 0.126\end{tabular} & \begin{tabular}[c]{@{}c@{}}0.855\\ 0.125\end{tabular} & \begin{tabular}[c]{@{}c@{}}0.832\\ 0.145\end{tabular} & \begin{tabular}[c]{@{}c@{}}0.853\\ 0.137\end{tabular} & \begin{tabular}[c]{@{}c@{}}0.914\\ 0.122\end{tabular} & \begin{tabular}[c]{@{}c@{}}0.956\\ 0.102\end{tabular}\\ \hline
\end{tabular}}
\end{table}

\begin{table}[!p]
\centering
\caption{
Averages and standard deviations of the trace correlation $R$ for the PSRFR, OPG, and MAVE methods for models {\bf [NE1]}, {\bf [NE2]}, and {\bf [NE3]} based on 1000 simulations with different sample sizes.}
\label{TNonE}
\setlength{\tabcolsep}{3pt} % Default value: 6pt
\renewcommand{\arraystretch}{1.4}
{\footnotesize
\begin{tabular}{ccccccccccc}
\hline
\multirow{2}{*}{\begin{tabular}[c]{@{}c@{}}\\ 
\end{tabular}} & & \multicolumn{3}{c}{{\bf Model [NE1]}} & \multicolumn{3}{c}{{\bf Model [NE2]}}  & \multicolumn{3}{c}{{\bf Model [NE3]}}\\ \cline{2-11}  & $n$ & 100 & 300 & 500& 100 & 300 & 500 & 100 & 300 & 500\\ \hline
PSRFR   & \begin{tabular}[c]{@{}c@{}}Average\\ SD\end{tabular} & \begin{tabular}[c]{@{}c@{}}0.778\\ 0.131\end{tabular} & \begin{tabular}[c]{@{}c@{}}0.911\\ 0.067\end{tabular} & \begin{tabular}[c]{@{}c@{}}0.945\\ 0.042\end{tabular} & \begin{tabular}[c]{@{}c@{}}0.894\\ 0.077\end{tabular} & \begin{tabular}[c]{@{}c@{}}0.956\\ 0.024\end{tabular} & \begin{tabular}[c]{@{}c@{}}0.974\\ 0.012\end{tabular} & \begin{tabular}[c]{@{}c@{}}0.913\\ 0.061\end{tabular} & \begin{tabular}[c]{@{}c@{}}0.968\\ 0.016\end{tabular} & \begin{tabular}[c]{@{}c@{}}0.982\\ 0.008\end{tabular} \\ 
OPG  & \begin{tabular}[c]{@{}c@{}}Average\\ SD\end{tabular} & \begin{tabular}[c]{@{}c@{}}0.880\\ 0.114\end{tabular} & \begin{tabular}[c]{@{}c@{}}0.989\\ 0.010\end{tabular} & \begin{tabular}[c]{@{}c@{}}0.995\\ 0.003\end{tabular} & \begin{tabular}[c]{@{}c@{}}0.987\\ 0.007\end{tabular} & \begin{tabular}[c]{@{}c@{}}0.997\\ 0.001\end{tabular} & \begin{tabular}[c]{@{}c@{}}0.999\\ 0.000\end{tabular} & \begin{tabular}[c]{@{}c@{}}0.964\\ 0.062\end{tabular} & \begin{tabular}[c]{@{}c@{}}0.995\\ 0.002\end{tabular} & \begin{tabular}[c]{@{}c@{}}0.997\\ 0.001\end{tabular} \\ 
MAVE  & \begin{tabular}[c]{@{}c@{}}Average\\ SD\end{tabular} & \begin{tabular}[c]{@{}c@{}}0.875\\ 0.117\end{tabular} & \begin{tabular}[c]{@{}c@{}}0.988\\ 0.008\end{tabular} & \begin{tabular}[c]{@{}c@{}}0.995\\ 0.002\end{tabular} & \begin{tabular}[c]{@{}c@{}}0.985\\ 0.024\end{tabular} & \begin{tabular}[c]{@{}c@{}}0.997\\ 0.002\end{tabular} & \begin{tabular}[c]{@{}c@{}}0.998\\ 0.001\end{tabular} & \begin{tabular}[c]{@{}c@{}}0.960\\ 0.089\end{tabular} & \begin{tabular}[c]{@{}c@{}}0.994\\ 0.003\end{tabular} & \begin{tabular}[c]{@{}c@{}}0.997\\ 0.001\end{tabular} \\ \hline
\end{tabular}}
\end{table}

\subsection{Effect of the dimension of predictor variables}
\label{sec3.3}

In the simulation studies presented in Sections \ref{sec3.1} and \ref{sec3.2}, the dimension of predictor variables is considered as $p = \dim(\bm{X}) = 10$ by following the classical works on SDR. In this subsection, we examine the performance of the proposed PSRFR method when the dimension of predictor variables $p$ is larger than 10. 

In this simulation study, we consider $p=30$ and $40$ with 
$\beta_{1}=(1,0,0,\dots,0)$ and $\beta_{2}=(0,1,0,\dots,0)$ under model {\bf [N4]} with multivariate normal distributed predictor variables and under model {\bf [NN3]} with multivariate Cauchy distribution, where the matrix 
$\Sigma_{norm}$ is a diagonal matrix with elements  $(1, 1, 1, 2, 2, 2, \dots,$\\$ 10, 10, 10)$ for $p = 30$ and  $(1, 1, 1, 1, 2, 2, 2, 2, \dots, 10, 10, 10, 10)$ for $p = 40$, and the matrix and $\Sigma_{ellp}$ 
is a diagonal matrix with elements $(1, 1, 1, 6, 6, 6, 11, 11, 11, \dots,$\\$ 46, 46, 46)$ for $p = 30$ and    $(1, 1, 1, 1, 6, 6, 6, 6, \dots,$ \\ $ 46, 46, 46, 46)$ for $p = 40$. 
We compare the performance of the proposed PSRFR method with the SIR and IHT methods when $p = 30$ and 40. Table \ref{DimP} presents the averages and standard deviations of the trace correlation $R$ for PSRFR, SIR, and IHT methods based on 1000 simulations.

\begin{table}[]
\centering
	\caption{
Averages and standard deviations of the trace correlation $R$ for the PSRFR method for model {\bf [N4]} with multivariate normally distributed predicted variables, model {\bf [NN3]}
with multivariate Cauchy distributed predicted variables based on 1000 simulations with different sample sizes.}
\label{DimP}
        \setlength{\tabcolsep}{5.5pt} % Default value: 6pt
        \renewcommand{\arraystretch}{1.1}
{\footnotesize
\begin{tabular}{cccllcll}
\hline
\multicolumn{2}{c}{\multirow{2}{*}{$p=30$}} & \multicolumn{3}{c}{Normal}  & \multicolumn{3}{c}{Cauchy}  \\ \cline{3-8}
\multicolumn{2}{c}{}  & \multicolumn{3}{c}{\bf Model  [N4]} & \multicolumn{3}{c}{\bf Model  [NN3]}\\ \hline
\multicolumn{1}{l}{} & \multicolumn{1}{l}{} & \multicolumn{1}{l}{$n=100$}& $n=300$  & $n=500$ & \multicolumn{1}{l}{$n=100$}  & $n=300$ & $n=500$ \\ \hline
PSRFR  & \begin{tabular}[c]{@{}c@{}}Average\\ SD\end{tabular} & \begin{tabular}[c]{@{}c@{}}0.510\\ 0.138\end{tabular}                     & \multicolumn{1}{c}{\begin{tabular}[c]{@{}c@{}}0.727\\ 0.136\end{tabular}} & \multicolumn{1}{c}{\begin{tabular}[c]{@{}c@{}}0.797\\ 0.123\end{tabular}} & \begin{tabular}[c]{@{}c@{}}0.699\\ 0.155\end{tabular}                     & \multicolumn{1}{c}{\begin{tabular}[c]{@{}c@{}}0.738\\ 0.160\end{tabular}} & \multicolumn{1}{c}{\begin{tabular}[c]{@{}c@{}}0.746\\ 0.158\end{tabular}} \\ 
SIR  & \begin{tabular}[c]{@{}c@{}}Average\\ SD\end{tabular} & \begin{tabular}[c]{@{}c@{}}0.271\\ 0.129\end{tabular}                     & \multicolumn{1}{c}{\begin{tabular}[c]{@{}c@{}}0.459\\ 0.125\end{tabular}} & \multicolumn{1}{c}{\begin{tabular}[c]{@{}c@{}}0.548\\ 0.115\end{tabular}} & \begin{tabular}[c]{@{}c@{}}0.435\\ 0.173\end{tabular}                     & \multicolumn{1}{c}{\begin{tabular}[c]{@{}c@{}}0.407\\ 0.164\end{tabular}} & \multicolumn{1}{c}{\begin{tabular}[c]{@{}c@{}}0.392\\ 0.166\end{tabular}} \\ 
IHT & \begin{tabular}[c]{@{}c@{}}Average\\ SD\end{tabular} & \begin{tabular}[c]{@{}c@{}}0.492\\ 0.110\end{tabular}    & \multicolumn{1}{c}{\begin{tabular}[c]{@{}c@{}}0.531\\ 0.051\end{tabular}} & \multicolumn{1}{c}{\begin{tabular}[c]{@{}c@{}}0.531\\ 0.052\end{tabular}} & \begin{tabular}[c]{@{}c@{}}0.324\\ 0.201\end{tabular}   & \multicolumn{1}{c}{\begin{tabular}[c]{@{}c@{}}0.337\\ 0.196\end{tabular}} & \multicolumn{1}{c}{\begin{tabular}[c]{@{}c@{}}0.325\\ 0.204\end{tabular}} \\ \hline
\multicolumn{2}{c}{$p=40$}  & \multicolumn{1}{c}{}  & \multicolumn{1}{c}{}   &  & \multicolumn{1}{c}{} & \multicolumn{1}{c}{}  \\ \hline
PSRFR   & \begin{tabular}[c]{@{}c@{}}Average\\ SD\end{tabular} & \multicolumn{1}{l}{\begin{tabular}[c]{@{}l@{}}0.364\\ 0.128\end{tabular}} & \begin{tabular}[c]{@{}l@{}}0.590\\ 0.140\end{tabular}   & \begin{tabular}[c]{@{}l@{}}0.688\\ 0.133\end{tabular}                     & \multicolumn{1}{l}{\begin{tabular}[c]{@{}l@{}}0.546\\ 0.157\end{tabular}} & \begin{tabular}[c]{@{}l@{}}0.624\\ 0.165\end{tabular}   & \begin{tabular}[c]{@{}l@{}}0.633\\ 0.166\end{tabular}  \\ 
SIR  & \begin{tabular}[c]{@{}c@{}}Average\\ SD\end{tabular} & \multicolumn{1}{l}{\begin{tabular}[c]{@{}l@{}}0.205\\ 0.107\end{tabular}} & \begin{tabular}[c]{@{}l@{}}0.369\\ 0.128\end{tabular}   & \begin{tabular}[c]{@{}l@{}}0.480\\ 0.111\end{tabular}   & \multicolumn{1}{l}{\begin{tabular}[c]{@{}l@{}}0.340\\ 0.155\end{tabular}} & \begin{tabular}[c]{@{}l@{}}0.325\\ 0.148\end{tabular}   & \begin{tabular}[c]{@{}l@{}}0.311\\ 0.146\end{tabular}     \\ 
IHT    & \begin{tabular}[c]{@{}c@{}}Average\\ SD\end{tabular} & \multicolumn{1}{l}{\begin{tabular}[c]{@{}l@{}}0.420\\ 0.174\end{tabular}} & \begin{tabular}[c]{@{}l@{}}0.512\\ 0.090\end{tabular}     & \begin{tabular}[c]{@{}l@{}}0.518\\ 0.071\end{tabular}   & \multicolumn{1}{l}{\begin{tabular}[c]{@{}l@{}}0.216\\ 0.216\end{tabular}} & \begin{tabular}[c]{@{}l@{}}0.200\\ 0.213\end{tabular}   & \begin{tabular}[c]{@{}l@{}}0.215\\ 0.215\end{tabular} \\ \hline
\end{tabular}}
\end{table}

From Table \ref{DimP}, we observe that the performances of PSRFR, SIR, and IHT methods depreciate when the dimension of $\bm{X}$ increases and the sample size decreases. This observation is consistent with the intuitive realization. Moreover, we observe that the PSRFR method still performs reasonably well ($R$ close to or greater than 0.7 in most cases) with the dimension of $\bm{X}$ being 30 and seems less sensitive to the increase in dimensionality when compared to the SIR and IHT methods. 

 \subsection{Effect of the general basis vectors and different noise levels}
\label{sec3.4}
In the simulation studies presented in Sections \ref{sec3.1} ,\ref{sec3.2} and \ref{sec3.3}, the number of basis vectors is considered as two and the noise level is 0.5. In this subsection, we examine the performance of the proposed PSRFR method when the basis vector becomes general under different noise levels.

In this simulation study, we consider $\sigma=2$ and $\sigma=4$ with 
$\beta_{1}=(1/\sqrt{2},1/\sqrt{2},0,$\\$\dots,0)$, $\beta_{2}=(1/\sqrt{2},-1/\sqrt{2},0,\dots,0)$, $\beta_{3}=(0,0,1/\sqrt{2},1/\sqrt{2},$ $0,\dots,0)$ and $\beta_{4}=(0,0,1/\sqrt{2},-1/\sqrt{2},0,\dots,0)$ under following model with $10$-dimensional multivariate normal distributed predictor variables in Sections \ref{sec3.1}. 
\begin{itemize}
    \item {\bf Model: }
    \begin{equation*}
    Y=\mathrm{sin}\left(\beta_{1}^\top\bm{X}+4\right)+\exp\left(\beta_{2}^\top\bm{X}\right )+\left(\beta_{3}^\top\bm{X}\right)^2+\left|\beta_{4}^\top\bm{X}\right| +\sigma\varepsilon.
    \end{equation*}
\end{itemize}

We compare the performance of the proposed PSRFR method with the PHD, SIR, SAVE, IHT, OPG and MAVE methods. Table \ref{Diffbeta} presents the averages and standard deviations of the trace correlation $R$ based on 1000 simulations.

\begin{table}[!ht]
\centering
	\caption{Averages and standard deviations of the trace correlation $R$ for the PSRFR, PHD, SIR, SAVE, IRE, IHT, OPG and MAVE methods for Model  based on 1000 simulations with different sample sizes.}
 \label{Diffbeta}
        \setlength{\tabcolsep}{4pt} % Default value: 6pt
\renewcommand{\arraystretch}{1.2}
{\footnotesize
\begin{tabular}{clcccccccc}
\hline
\multicolumn{10}{c}{$\sigma=2$} \\ \hline
\multicolumn{3}{c}{Method} &
  PSRFR & PHD & SIR & SAVE & IHT & OPG & MAVE \\ \hline
\multicolumn{2}{c}{$n=100$} &
  \begin{tabular}[c]{@{}c@{}}Average\\ SD\end{tabular} &
  \begin{tabular}[c]{@{}c@{}}0.8817\\ 0.0612\end{tabular} &
  \begin{tabular}[c]{@{}c@{}}0.7292\\ 0.0737\end{tabular} &
  \begin{tabular}[c]{@{}c@{}}0.6233\\ 0.0802\end{tabular} &
  \begin{tabular}[c]{@{}c@{}}0.7007\\ 0.0863\end{tabular} &
  \begin{tabular}[c]{@{}c@{}}0.5912\\ 0.1054\end{tabular} &
  \begin{tabular}[c]{@{}c@{}}0.7938\\ 0.0703\end{tabular} &
  \begin{tabular}[c]{@{}c@{}}0.7914\\ 0.0737\end{tabular} \\
\multicolumn{2}{c}{$n=300$} &
  \begin{tabular}[c]{@{}c@{}}Average\\ SD\end{tabular} &
  \begin{tabular}[c]{@{}c@{}}0.9472\\ 0.0356\end{tabular} &
  \begin{tabular}[c]{@{}c@{}}0.7803\\ 0.0836\end{tabular} &
  \begin{tabular}[c]{@{}c@{}}0.6807\\ 0.0895\end{tabular} &
  \begin{tabular}[c]{@{}c@{}}0.7734\\ 0.0733\end{tabular} &
  \begin{tabular}[c]{@{}c@{}}0.6280\\ 0.0808\end{tabular} &
  \begin{tabular}[c]{@{}c@{}}0.8610\\ 0.0787\end{tabular} &
  \begin{tabular}[c]{@{}c@{}}0.8562\\ 0.0799\end{tabular} \\
\multicolumn{2}{c}{$n=500$} &
  \begin{tabular}[c]{@{}c@{}}Average\\ SD\end{tabular} &
  \begin{tabular}[c]{@{}c@{}}0.9677\\ 0.0161\end{tabular} &
  \begin{tabular}[c]{@{}c@{}}0.8257\\ 0.0781\end{tabular} &
  \begin{tabular}[c]{@{}c@{}}0.6881\\ 0.0944\end{tabular} &
  \begin{tabular}[c]{@{}c@{}}0.8208\\ 0.0714\end{tabular} &
  \begin{tabular}[c]{@{}c@{}}0.6290\\ 0.0802\end{tabular} &
  \begin{tabular}[c]{@{}c@{}}0.8832\\ 0.0930\end{tabular} &
  \begin{tabular}[c]{@{}c@{}}0.8720\\ 0.0890\end{tabular} \\ \hline
\multicolumn{10}{c}{$\sigma=4$} \\ \hline
\multicolumn{2}{c}{$n=100$} &
  \begin{tabular}[c]{@{}c@{}}Average\\ SD\end{tabular} &
  \begin{tabular}[c]{@{}c@{}}0.8812\\ 0.0576\end{tabular} &
  \begin{tabular}[c]{@{}c@{}}0.7280\\ 0.0824\end{tabular} &
  \begin{tabular}[c]{@{}c@{}}0.6984\\ 0.0750\end{tabular} &
  \begin{tabular}[c]{@{}c@{}}0.6895\\ 0.0839\end{tabular} &
  \begin{tabular}[c]{@{}c@{}}0.5787\\ 0.0987\end{tabular} &
  \begin{tabular}[c]{@{}c@{}}0.7670\\ 0.0645\end{tabular} &
  \begin{tabular}[c]{@{}c@{}}0.7679\\ 0.0710\end{tabular} \\
\multicolumn{2}{c}{$n=300$} &
  \begin{tabular}[c]{@{}c@{}}Average\\ SD\end{tabular} &
  \begin{tabular}[c]{@{}c@{}}0.9475\\ 0.0340\end{tabular} &
  \begin{tabular}[c]{@{}c@{}}0.7625\\ 0.0898\end{tabular} &
  \begin{tabular}[c]{@{}c@{}}0.6858\\ 0.0889\end{tabular} &
  \begin{tabular}[c]{@{}c@{}}0.7546\\ 0.0825\end{tabular} &
  \begin{tabular}[c]{@{}c@{}}0.6116\\ 0.0923\end{tabular} &
  \begin{tabular}[c]{@{}c@{}}0.8222\\ 0.0768\end{tabular} &
  \begin{tabular}[c]{@{}c@{}}0.8244\\ 0.0738\end{tabular} \\
\multicolumn{2}{c}{$n=500$} &
  \begin{tabular}[c]{@{}c@{}}Average\\ SD\end{tabular} &
  \begin{tabular}[c]{@{}c@{}}0.9684\\ 0.0158\end{tabular} &
  \begin{tabular}[c]{@{}c@{}}0.7981\\ 0.0783\end{tabular} &
  \begin{tabular}[c]{@{}c@{}}0.6848\\ 0.0849\end{tabular} &
  \begin{tabular}[c]{@{}c@{}}0.7787\\ 0.0831\end{tabular} &
  \begin{tabular}[c]{@{}c@{}}0.6307\\ 0.0854\end{tabular} &
  \begin{tabular}[c]{@{}c@{}}0.9463\\ 0.0756\end{tabular} &
  \begin{tabular}[c]{@{}c@{}}0.8485\\ 0.0712\end{tabular} \\ \hline
\end{tabular}
}
\end{table}

From Table \ref{Diffbeta}, we observe that the more general basis vectors with different noise levels do not have much effect on the performance of the PSRFR method.

 In summary, as demonstrated by the simulation results in Sections \ref{sec3.1}--\ref{sec3.4}, the proposed PSRFR method exhibits promising performance, particularly evident when the variances of individual components diverge, and the tails of predictor variable distributions become heavier. Notably, PSRFR maintains its robustness and reliability even in scenarios where predictor variable distributions deviate from the elliptical distribution assumption and for large dimensions of the predictor variables. This versatility renders PSRFR applicable to a broader spectrum of real-world data analyses, enhancing its practical utility.

\section{Real Data Analysis}
\label{RealData}
The Wine Quality dataset presented in \cite{CORTEZ2009547} is a publicly available data set that contains two sub-datasets: the red wine and white wine data sets. 
The response variable is the quality of wine (QL), and the 11 predictor variables are: fixed acidity (FA); volatile acidity (VA); citric acid (CA); residual sugar (RS);  chlorides (CL); free sulfur dioxide (FSD); total sulfur dioxide (TSD);  density (DS); pH value (PH);  sulphates (SP); and alcohol level (AH). \citet{CORTEZ2009547} studied the relative importance of 11 predictor variables for wine quality using the support vector machine approach and pointed out that a regression approach on the two sub-datasets can be used. Here, we consider the regression approach and apply the proposed PSRFR method to obtain the relative importance with the first 1599 and 800 observations in the red and white wine sub-datasets, respectively.  
And  we compare to the result obtained by the support vector machine approach.

\begin{figure}[!h]
\centering
\includegraphics[width=0.65\textwidth]{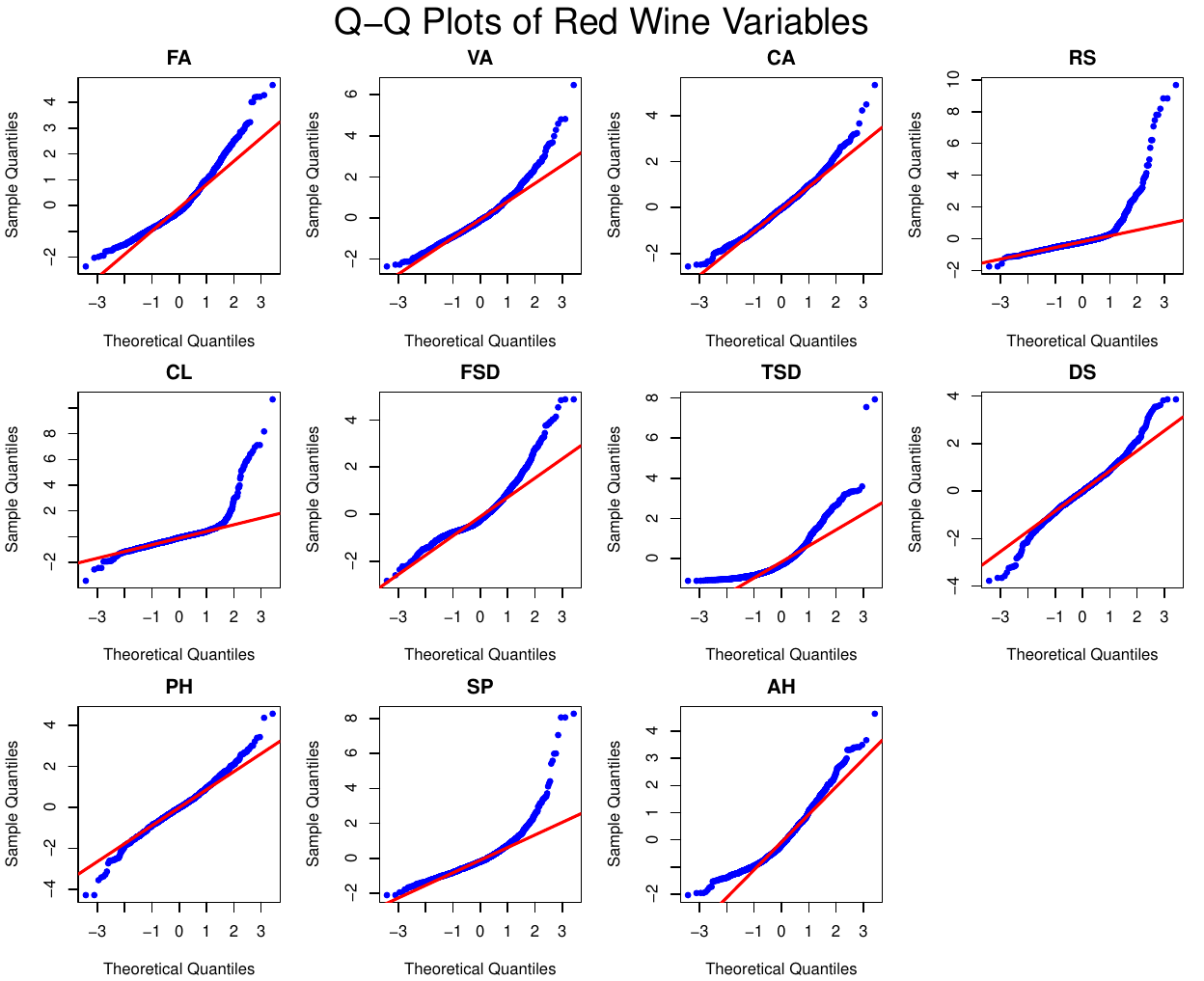}
    \caption{The normal quantile-quantile plots of the 1599 observations for different variables in the red wine data set.}
\label{fig:qqred}
\end{figure}

\begin{figure}[!h]
    \centering
    \includegraphics[width=0.65\textwidth]{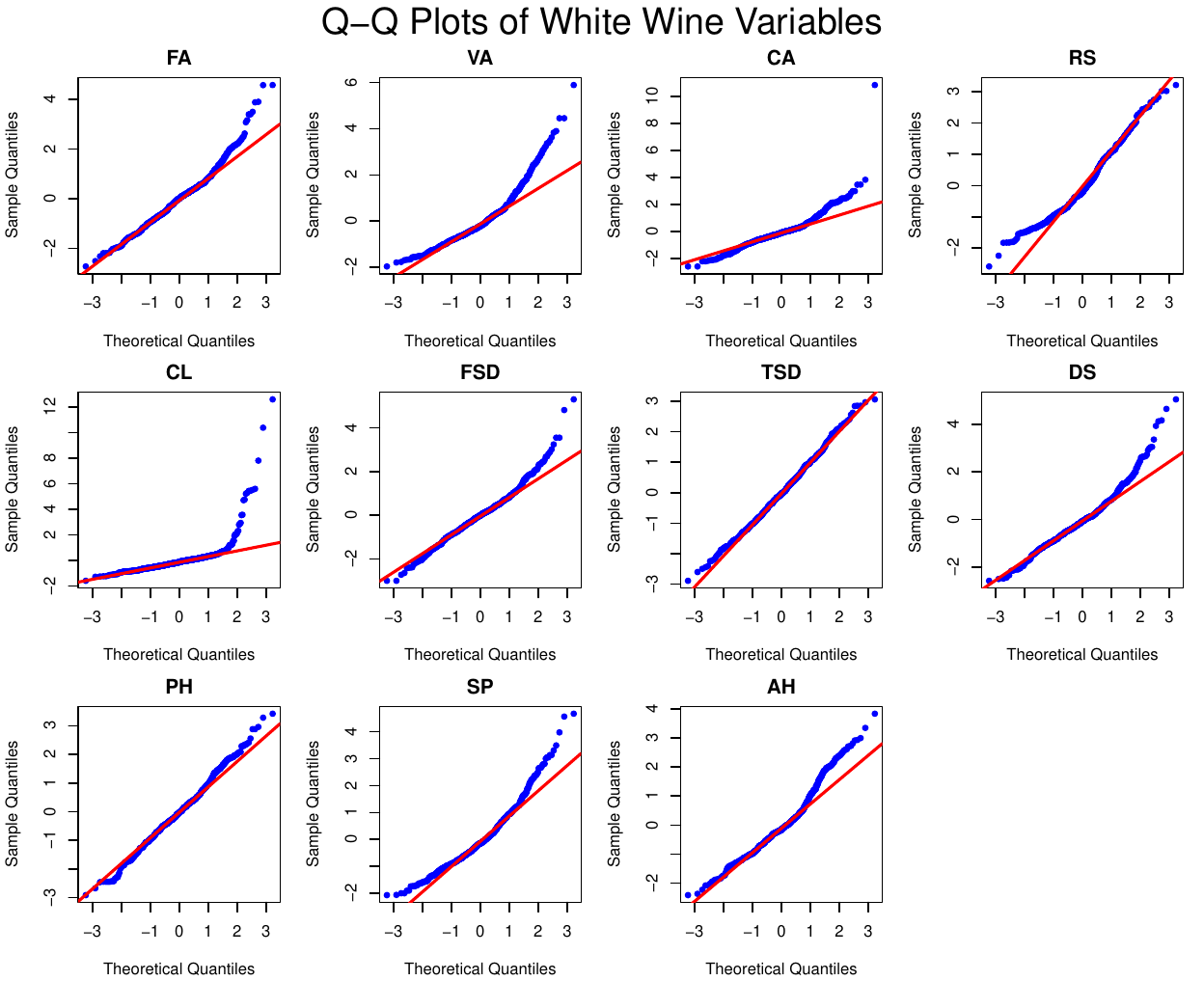}
    \caption{The normal quantile-quantile plots of the 800 observations for different variables in the white wine data set}
    \label{fig:qqwhite}
\end{figure}

First, we assess the normality of the 11 predictor variables using the hypothesis testing approach and graphical approach, namely the Shapiro-Wilk test and normal quantile-quantile (Q-Q) plot, after standardizing the data. 
Table \ref{TReal} presents the Shapiro-Wilk statistics for predictor variables in the red wine and white wine data sets. The normal Q-Q plots are depicted in Figure \ref{fig:qqred} and \ref{fig:qqwhite} for the red wine and white wine data sets, respectively. From Table \ref{TReal} and the normal Q-Q plots in Figure \ref{fig:qqred} and \ref{fig:qqwhite}, except for variables TSD and PH in the white wine data set, all the other variables in both data sets are not likely to follow a normal distribution.  

\begin{figure}[!h]
    \centering
\includegraphics[width=0.7\textwidth]{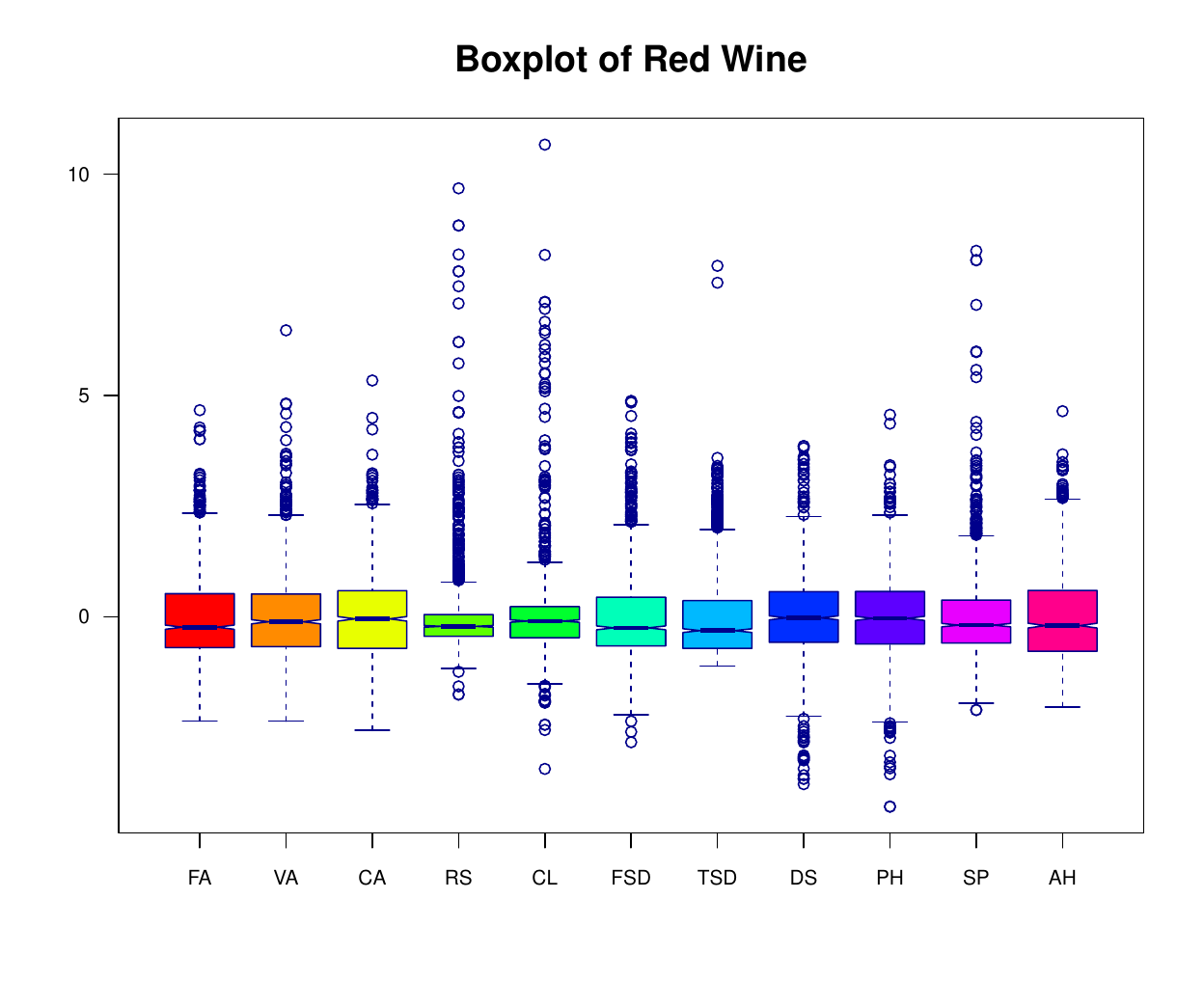}
\caption{The comparative boxplot based on the 1599 observations for the 11 predictor variables (after standardization) in the red wine data set.}
\label{fig:boxred}
\end{figure}

\begin{figure}[!h]
    \centering
\includegraphics[width=0.65\textwidth]{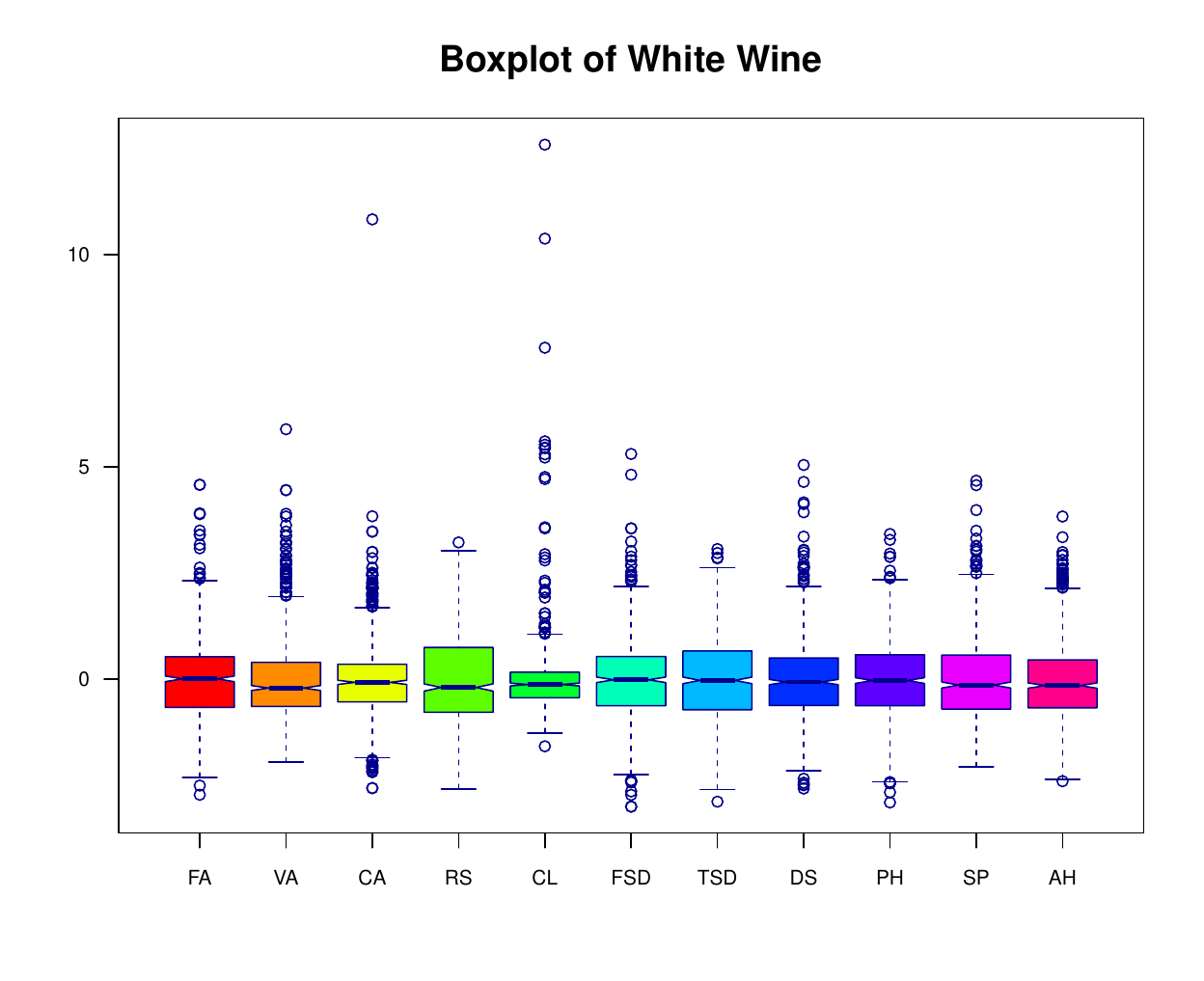}
    \caption{The comparative boxplot based on the 800 observations for the 11 predictor variables (after standardization) in the white wine data set.}
\label{fig:boxwhite}
\end{figure}

To assess the symmetry of the distributions of these 11 predictor variables in the red wine and white wine data sets, we provide the comparative boxplots in Figure \ref{fig:boxred} and \ref{fig:boxwhite}. From Figure \ref{fig:boxred} and \ref{fig:boxwhite}, we observe that 
the variables in the red wine and white wine data sets are asymmetric and have heavy-tailed characteristics.

Considering that the PSRFR method performs well when the variances of the predictors are significantly different from each other, we diagonalize the data by performing an eigen-decomposition on the sample covariance matrix and multiply the centered data by the corresponding eigenvectors. Then, we apply the PSRFR method to address the regression problem at hand. Once the PSRFR estimator is obtained, the next step involves determining the dimension of the central subspace. We consider evaluating the proportion that an eigenvalue accounts for all the eigenvalues to determine the dimension of the central space, which is similar to the way of determining the number of principal components in PCA. The results show that the proportions of the first eigenvalues in the red wine and white wine data sets account for all the eigenvalues 
are 0.9995 and 0.9997, respectively, which indicate that the dimension of central space can be considered as $1$. 

Next, our objective is to estimate the basis $\hat{\beta}_{1}$ of the central space. To account for relative importance, we take the absolute value of each component in $\hat{\beta}_{1}$ and arrange them in descending order, from largest to smallest. In the red wine data set, the relative importance in descending order is AH, PH, SP, DS, FSD, TSD, CL, RS, CA, VA, and FA. In the white wine data set, the relative importance in descending order is AH, SP, TSD, DS, CL, FSD, RS, PH, CA, FA, VA. These results align with the findings in \cite{CORTEZ2009547}, in which PH, SP, and AH are also identified as important variables in the red wine data set, while RS and CA were relatively unimportant. In the white wine data set, SP and AH are relatively important, while FA and PH are relatively unimportant. 

We observe that both SP and AH are highly important variables in both red wine and white wine data sets. \cite{CORTEZ2009547} provided a physiological explanation about this and suggested that an increase in sulfates may be associated with fermenting nutrition, which plays a crucial role in enhancing the wine aroma. The significance of AH in wine is evident. Furthermore, the importance of pH value (PH) in red wine surpasses that in white wine. Although SP and AH are consistently identified as important variables, it is noteworthy that SP holds the highest importance in \cite{CORTEZ2009547}, and AH emerges as the most important variable in our findings. \cite{CORTEZ2009547} suggests that an increase in alcohol content tends to result in higher-quality wine. 
Furthermore, considering that AH holds the highest significance in our results and DS is influenced by the proportion of AH and sugar content, it can be inferred that DS may have a greater importance than initially suggested in the study by \cite{CORTEZ2009547}. 

\begin{table}[H]
\centering
  \caption{The test statistics and $p$-values of Shapiro-Wilk normality tests for the 11 variables in the red wine and white wine data sets.}
  \label{TReal}
  \setlength{\tabcolsep}{3pt} % Default value: 6pt
  \renewcommand{\arraystretch}{1.2}
\begin{threeparttable}
{\tiny
\begin{tabular}{ccccccccccccc}
\hline
 & & FA & VA & CA & RS & CL & FSD & TSD & DS & PH & SP & AH \\ \hline
\multirow{2}{*}{Red} & Test Stat. & 0.941 & 0.953 & 0.984 & 0.601 & 0.680 & 0.908 & 0.833 & 0.983 & 0.990 & 0.838 & 0.944 \\
& $p$-value & $<0.001$ & $<0.001$ & $<0.001$ & $<0.001$ & $<0.001$ & $<0.001$ & $<0.001$ & $<0.001$ & $<0.001$ & $<0.001$ & $<0.001$ \\ \hline
\multirow{2}{*}{White} & Test Stat. & 0.974 & 0.908 & 0.889 & 0.968 & 0.531 & 0.978 & 0.997 & 0.961 & 0.995 & 0.951 & 0.962 \\
 & $p$-value & $<0.001$ & $<0.001$ & $<0.001$ & $<0.001$ & $<0.001$ & $<0.001$ & $<0.1$ & $<0.001$ & $<0.05$ & $<0.001$ & $<0.001$\\ \hline
\end{tabular}}
        
\end{threeparttable}
\end{table}

\section{Concluding Remarks}
\label{Concluding}
In this paper, we propose a principal square response forward regression (PSRFR) method, a novel approach for dimension reduction tailored for high-dimensional, elliptically distributed data. Drawing inspiration from the Ordinary Least Squares (OLS) method, PSRFR is devised to handle the complexities of such datasets effectively.

The core principle of PSRFR lies in leveraging the amalgamated information from both predictor and response variables, which typically concentrates around a central subspace. Unlike the OLS method, which tends to recover only a single dimension, PSRFR aims at capturing a comprehensive estimate of this central subspace. The PSRFR method achieves this by minimizing the distance between data points and the central subspace, thus identifying multiple central subspace directions, surpassing the capabilities of methods like PHD and IHT methods when the predictor variables follow an elliptical distribution. Moreover, this paper presents a fundamental theorem affirming the efficacy of PSRFR in achieving substantial dimension reduction. Additionally, we provide a simple algorithm for implementing PSRFR. We delve into the asymptotic behavior of the PSRFR estimator, elucidating its convergence rate in high-dimensional scenarios.

Our simulation results underscore the superiority of PSRFR in enhancing estimation accuracy for elliptically distributed data with varying component variances. This improvement is facilitated through a simple data transformation process. Overall, the proposed PSRFR method furnishes invaluable tools for dimension reduction and analysis of high-dimensional, elliptically distributed data, exhibiting resilience even in the face of deviations from the elliptical distribution. 

Note that determining the dimension of the central subspace is a critical issue, which we leave for further study.

\subsection*{Competing interests}
The authors declare there are no conflict of interests.

\subsection*{Funding}
The research is supported by the National Natural Science Foundation of China (No. 12371263).

\appendix
\section{Appendix}
\subsection{Proof of Lemma \ref{Lemma1}}

%\begin{proof}
%Here, we consider proofing Lemma \ref{Lemma1} from a different perspective. 
For  $\text{B}=\left(\beta_{1},\ldots,\beta_{k}\right)$, without loss of generality, we consider $\mathrm{E}(\bm{X}) = 0$ (otherwise, we can consider transforming  $\bm{X}$ by $\bm{X} - \mathrm{E}(\bm{X})$). \\

\noindent 
{\bf Case 1:  $\Sigma_{X}=\textup{I}_{p}$.} In this case, we have 
\begin{align*}
\mathrm{E}(Y\bm{X})&=\mathrm{E}\Big\{\mathrm{E}\left(Y\bm{X}\mid \bm{X}\right)\Big\}\\
&=\mathrm{E}\Big\{\bm{X}\cdot \mathrm{E}(Y\mid \bm{X})\Big\}
=\mathrm{E}\Big\{h\left(\beta_{1}^\top\bm{X},\cdots,\beta_{k}^\top\bm{X}\right)\cdot \bm{X} \Big\}.
\end{align*}
Let $\text{C}$ be the $p\times p$ orthogonal matrix with the first $k$ rows be $\beta_{i}^\top,\ i\in\{1,\ldots,k\}$. 
Then, define 
\begin{equation*}
\text{C} =(\beta_{1},\dots,\beta_{k},\alpha_{k+1},\dots,\alpha_{p})^\top\equiv \left(\text{B}, \text{A} \right)^\top,  
\end{equation*}
and let
\begin{align*}
\bm{W}&=\text{C}\bm{X}=
(\beta_{1}^\top\bm{X},\cdots,\beta_{k}^\top\bm{X}, \alpha_{k+1}^\top\bm{X},\cdots,\alpha_{p}^\top\bm{X})^\top\\
&=(w_{1},\cdots,w_{k},w_{k+1},\cdots,w_{p})^\top.
\end{align*}
Hence,
\begin{align*}
E(Y\bm{X})&=E\left[h\left(\beta_{1}^\top \bm{X},\cdots,\beta_{k}^\top \bm{X}\right)\cdot \bm{X}\right]\\
&=C^\top E\left[h\left(\beta_{1}^\top \bm{X},\dots,\beta_{k}^\top \bm{X}\right)\cdot C\bm{X}\right]\\
&=C^\top E\Big[h\left(w_{1},\dots,w_{k}\right)\cdot \bm{W}\Big].
\end{align*}
Note that $\bm{W} = \text{C}\bm{X}$ also follows elliptical distributions with with mean $\mathrm{E}(\bm{W}) = 0$ and variance-covariance matrix $\text{C}\textup{I}_{p}\text{C}^\top=\textup{I}_{p}$, where $\textup{I}_{p}$ is a $p$-dimensional identity matrix. By Theorem 7 in \citet{frahm2004generalized}, we can obtain 
\begin{equation*}
	\mathrm{E}(w_{j}\mid w_{1}, \ldots,w_{k})=0, \ j\in\{k+1,\ldots,p\},
\end{equation*}
hence,
\begin{align*}
\mathrm{E}&\Big\{h\left(w_{1},\ldots,w_{k}\right) \ldots w_{j}\Big\}=\mathrm{E}\Big \{ \mathrm{E}\Big ( h\left(w_{1},\ldots,w_{k}\right) \cdot w_{j} \mid w_{1},\ldots, w_{k} \Big ) \Big \}\\
&=\mathrm{E}\Big \{ h\left(w_{1},\ldots,w_{k}\right) \cdot \mathrm{E}(w_{j}\mid w_{1}, \ldots,w_{k}) \Big \}=0,\; j\in\{k+1,\ldots,p\}.
\end{align*}
Then,
\begin{align*}
    \mathrm{E}&(Y\bm{X})=\text{C}^\top\mathrm{E}\Big\{h\left(w_{1},\ldots,w_{k}\right)\cdot \bm{W}\Big\} \\
 &=\text{C}^\top\left[\mathrm{E}\Big\{h\left(w_{1},\ldots,w_{k}\right)\cdot w_{1}\Big\},\dots,\mathrm{E}\Big\{h\left(w_{1},\ldots,w_{k}\right)\cdot w_{k}\Big\},0,\dots,0\right]^\top\\
 %& \begin{pmatrix}
%\mathrm{E}\Big[h\left(w_{1},\ldots,w_{k}\right)\cdot w_{1}\Big],
%\ldots, 
%\mathrm{E}\Big[h\left(w_{1},\ldots,w_{k}\right)\cdot w_{k}\Big], 
%0,
%\ldots,
%0
%\end{pmatrix}^\top \\
&=\sum_{i=1}^{k}\beta_{i}\mathrm{E}\Big\{h\left(w_{1},\ldots,w_{k}\right)\cdot w_{i}\Big\}
=\sum_{i=1}^{k}\textup{I}_{p}\beta_{i}\mathrm{E}\Big\{h\left(w_{1},\ldots,w_{k}\right)\cdot w_{i}\Big\}=\Sigma_{X}\text{B}\Lambda,
\end{align*}
where $\Lambda=(\lambda_{1},\ldots,\lambda_{k})^\top$, 
$ \lambda_{i}=\mathrm{E}\Big\{h\left(w_{1},\ldots,w_{k}\right)\cdot w_{i}\Big\}.$\\

\noindent 
{\bf Case 2: $\Sigma_{X}=\Sigma$.} Let $\bm{X}^{*}=\Sigma^{-1/2}\bm{X}$, then $\bm{X}^{*}$ follows  elliptical distribution with mean $0$ and variance-covariance matrix $\textup{I}_{p}$. According to Eq. \eqref{Y||E(y|X)}, we have
\begin{equation*}
   \mathrm{E}(Y\mid \bm{X})=\mathrm{E}(Y\mid \Sigma^{1/2}\bm{X}^{*})=h\left(\bm{X}^{*^\top}\Sigma^{1/2}\beta_{1},\cdots,\bm{X}^{*^\top}\Sigma^{1/2}\beta_{k}\right).
\end{equation*}
Hence,
\begin{equation*}
\mathrm{E}(Y\bm{X})=\Sigma^{1/2}\mathrm{E}(Y\bm{X}^{*})
=\Sigma^{1/2}\Sigma_{X^{*}}\Sigma^{1/2}\text{B}\Lambda= \Sigma_{X}\text{B}\Lambda.
\end{equation*}
%\end{proof}

\subsection{Proof of Theorem \ref{thm1}}

Without loss of generality, we consider  $\mathrm{E}(\bm{X})=0$.\\ 

\noindent 
{\bf Case 1:  $\Sigma_{X}=\textup{I}_{p}$.} In this case, 
\begin{equation*}
    \mathrm{E}(Y^{2}\mid \bm{X})=H\left(\beta_{1}^\top\bm{X},\ldots,\beta_{k}^\top\bm{X}\right),
\end{equation*}
then,
\begin{align*}
\mathrm{E}\Big(Y^{2}\bm{X}\bm{X}^\top\Big)&=\mathrm{E}\left \{ \mathrm{E}(Y^{2}\bm{X}\bm{X}^\top\mid \bm{X}) \right \}\\
&=\mathrm{E}\left \{ \bm{X}\bm{X}^\top\mathrm{E}(Y^{2}\mid \bm{X}) \right \}=\mathrm{E}\left \{ \bm{X}\bm{X}^\top H\left(\beta_{1}^\top\bm{X},\ldots,\beta_{k}^\top\bm{X}\right) \right \}.
\end{align*}
Similarly, let $\text{C}$ be the orthogonal matrix with the first $k$ rows as $\beta_{i}^\top,\ i\in\{1,\ldots,k\}$, then we define 
\begin{equation*}
\text{C}=(\beta_{1},\dots,\beta_{k},\alpha_{k+1},\dots,\alpha_{p})^\top\equiv\left(\text{B}, \text{A} \right)^\top,  
\end{equation*}
and hence,
\begin{align*}
\mathrm{E}(Y^{2}\bm{X}\bm{X}^\top)&=\mathrm{E}\left\{H\left(\beta_{1}^\top\bm{X}, \ldots,\beta_{k}^\top\bm{X}\right)\cdot \bm{X}\bm{X}^\top\right\}\\
&=\text{C}^\top\mathrm{E}\left\{H\left(\beta_{1}^\top\bm{X},\ldots,\beta_{k}^\top\bm{X}\right)\cdot \text{C}\bm{X}\bm{X}^\top \text{C}^\top\right\}\text{C}\\
&=\text{C}^\top\mathrm{E}\Big\{H\left(w_{1},\dots,w_{k}\right)\cdot \bm{W}\bm{W}^\top\Big\}\text{C}.
\end{align*}
Note that $\bm{W}=\text{C}\bm{X}$ also follows elliptical distributions with with mean $\mathrm{E}(\bm{W})=0$ and variance-covariance matrix $\text{C}\text{I}_{p}\text{C}^\top=\textup{I}_{p}$. Let
\begin{equation*}
    \bm{W}=\left( w_{1},\ldots,w_{k},w_{k+1},\ldots,w_{p} \right)^\top=\Big(\bm{W}_{(1)}^\top,\bm{W}_{(2)}^\top\Big)^\top.
\end{equation*} 
By Corollary 8 in \citet{frahm2004generalized}, we can obtain
\begin{align*}
    \mathrm{E}\left ( \bm{W}_{(2)}\bm{W}_{(2)}^\top\mid \bm{W}_{(1)} \right )
 &=\mathrm{diag}\Big \{\mathrm{E}\left ( w_{k+1}^{2}\mid \bm{W}_{(1)} \right ),\dots,\mathrm{E}\left ( w_{p}^{2}\mid \bm{W}_{(1)} \right )\Big \}\text{I}_{p-k}\\
 &=\mathrm{E}\left ( w_{k+1}^{2}\mid \bm{W}_{(1)} \right )\text{I}_{p-k}
\end{align*}
because $\{w_{k+1},\dots,w_{p}\}$ have the same status. Therefore,
\begin{align*}
E\left(Y^{2}\bm{X}\bm{X}^\top \right)&=C^\top E\Big[H\left(w_{1},\dots,w_{k}\right)\cdot \bm{W}\bm{W}^\top \Big]C\\
 &=C^\top E\bigg[ E \Big[ H\left(w_{1},\dots,w_{k}\right)\cdot \bm{W}\bm{W}^\top \mid \bm{W}_{(1)}  \Big] \bigg]C\\
 &=C^\top E\Big [ H\left(w_{1},\dots,w_{k}\right)\cdot E\left(\bm{W}\bm{W}^\top \mid \bm{W}_{(1)}\right) \Big ]C\\
 &=C^\top \begin{pmatrix}
\Gamma_{1} & 0\\ 
 0& \Gamma_{2}
\end{pmatrix}C\\
&=B\Gamma _{1}B^\top +A\Gamma _{2}A^\top , 
\end{align*}
where 
\begin{equation*}
    \Gamma_{1}=\begin{pmatrix}
\mathrm{E}\Big\{   H\left(w_{1},\dots,w_{k} \right )w_{1}^{2} \Big\}& \cdots &  \mathrm{E}\Big\{  H\left(w_{1},\dots,w_{k}\right) w_{1}w_{k} \Big\}\\ 
 \vdots & \ddots & \vdots \\ 
\mathrm{E}\Big\{  H\left(w_{1},\dots,w_{k}\right) w_{k}w_{1} \Big\}& \cdots & \mathrm{E}\Big\{  H\left(w_{1},\dots,w_{k}\right) w_{k}^{2} \Big\}
\end{pmatrix}_{k\times k},
\end{equation*}

\begin{align*}
 \Gamma_{2}&=\begin{pmatrix}
\mathrm{E}\Big\{  \mathrm{E}\left (H\cdot w_{k+1}^{2}\mid \bm{W}_{(1)} \right ) \Big\}& \cdots & 0\\ 
 \vdots &  \ddots & \vdots \\ 
 0& \cdots & \mathrm{E}\Big\{  \mathrm{E}\left (H\cdot w_{p}^{2}\mid \bm{W}_{(1)} \right ) \Big\}
\end{pmatrix}_{(p-k)\times (p-k)}\\
&=\mathrm{E}\Big\{  \mathrm{E}\left (H\left(w_{1},\dots,w_{k}\right)\cdot w_{k+1}^{2}\mid \bm{W}_{(1)} \right )\Big\}\text{I}_{p-k}\\
&=\mathrm{E}\left\{ H\left(w_{1},\dots,w_{k}\right)\cdot w_{k+1}^{2}\right\}\text{I}_{p-k}\\
&=\mathrm{E}\left\{ \mathrm{E}\left(Y^{2}\mid \bm{X}\right)\cdot w_{k+1}^{2}\right\}\text{I}_{p-k}\\
&=\mathrm{E}\left(Y^{2}\cdot w_{k+1}^{2}\right)\text{I}_{p-k}=\mathrm{E}\left(\text{G}\right)\text{I}_{p-k}.
\end{align*}
Next, note that
$\mathrm{E}\{Y^{2}(\beta^\top\bm{X})^{2}\}=\beta^\top\mathrm{E}(Y^{2}\bm{X}\bm{X}^\top)\beta=\beta^\top\text{B}\Gamma_{1}\text{B}^\top\beta$ is the eigenvalue of $\Gamma_{1}$,
and $\mathrm{E}\{Y^{2}(\alpha^\top\bm{X})^{2}\}$ is the eigenvalue of $\Gamma_{2}$. Assumption \ref{Ass2} assures that the eigenvalue of $\Gamma_{1}$ is larger, hence, the first $k$ eigenvectors corresponding to the first $k$ eigenvalues of $\mathrm{E}(Y^{2}\bm{X}\bm{X}^\top)$ are the basis of $\boldsymbol{S}_{\mathrm{E}\left({Y\mid \bm{X}}\right)}$.

Furthermore, $\mathrm{E}(Y^{2}\bm{X}\bm{X}^\top)$ can be rewritten as
\begin{align*}
\mathrm{E}\left(Y^{2}\bm{X}\bm{X}^\top\right)&=\text{B}\Gamma _{1}\text{B}^\top+\text{A}\Gamma _{2}\text{A}^\top\\
&=\text{B}\Gamma _{1}\text{B}^\top +\mathrm{E}\left(\text{G}\right)\text{AA}^\top  +\mathrm{E}\left(\text{G}\right)\text{BB}^\top 
-\mathrm{E}\left(\text{G}\right)\text{BB}^\top \\
&=\text{B}\Big\{ \Gamma _{1}-\mathrm{E}\left(\text{G}\right)\textup{I}_{k} \Big\}\text{B}^\top +\mathrm{E}\left(\text{G}\right)\textup{I}_{p},
\end{align*}
and we can obtain
\begin{equation*}
    \mathrm{E}\left(Y^{2}\bm{X}\bm{X}^\top\right)-\mathrm{E}\left(\text{G}\right)\textup{I}_{p}=\text{B}\left \{ \Gamma _{1}-\mathrm{E}\left(\text{G}\right)\textup{I}_{k} \right \}\text{B}^\top \equiv \text{BM}.
\end{equation*}
Because $\Gamma _{1}-\mathrm{E}\left(\text{G}\right)\textup{I}_{k}$ is a positive definite matrix by Assumption \ref{Ass2}, 
$\mathrm{rank}\text{(M)}=k$ and $\mathrm{rank}\left(\text{BM}\right)=\mathrm{rank}(\text{B})$ according to the results in  Section A4.4 of \cite{seber2003linear}, i.e., $\mathrm{E}(Y^{2}\bm{X}\bm{X}^\top)-\mathrm{E}\left(\text{G}\right)\textup{I}_{p}$ is contained in the linear subspace Spanned by the basis matrix $\text{B}$.\\

\noindent 
{\bf Case 2: $\Sigma_{X}=\Sigma$.} Let $\bm{X}^{*}=\Sigma^{-1/2}\bm{X}$, where $\bm{X}^{*}$ follows elliptical distribution with mean $0$ and variance-covariance matrix $\textup{I}_{p}$, then
\begin{equation*}
    \mathrm{E}(Y^{2}\mid \bm{X})=\mathrm{E}(Y^{2}\mid \Sigma^{1/2}\bm{X}^{*})=H\left(\bm{X}^{*^\top }\Sigma^{1/2}\beta_{1},\cdots,\bm{X}^{*^\top }\Sigma^{1/2}\beta_{k}\right), 
\end{equation*}
and 
\begin{align*}
    \mathrm{E}(Y^{2}\bm{X}\bm{X}^\top )&=\Sigma^{1/2}\mathrm{E}(Y^{2}\bm{X}^{*}\bm{X}^{*^\top })\Sigma^{1/2}\\
    &=\Sigma^{1/2}\Sigma^{1/2}\text{B}\Gamma _{1}\text{B}^\top \Sigma^{1/2}\Sigma^{1/2}+\Sigma^{1/2}\Sigma^{1/2}\text{A}\Gamma _{1}\text{A}^\top \Sigma^{1/2}\Sigma^{1/2}\\
    &=\Sigma \text{B}\Gamma _{1}\text{B}^\top \Sigma+\Sigma \text{A}\Gamma _{1}\text{A}^\top \Sigma\\
    &=\Sigma_{X}\text{B}\left \{ \Gamma _{1}-\mathrm{E}\left(\text{G}\right)\textup{I}_{k} \right \}\text{B}^\top \Sigma_{X}+\Sigma_{X}\mathrm{E}\left(\text{G}\right)\textup{I}_{p}.
\end{align*}
We can obtain
\begin{equation*}
    \mathrm{E}(Y^{2}\bm{X}\bm{X}^\top )-\mathrm{E}\left(\text{G}\right)\textup{I}_{p}\Sigma_{X}=\Sigma_{X}\text{B}\left \{ \Gamma _{1}-\mathrm{E}\left(\text{G}\right)\textup{I}_{k} \right \}\text{B}^\top \Sigma_{X}.
\end{equation*}
Similarily, the eigenvalue of $\Gamma_{1}$ is larger and $\Gamma _{1}-\mathrm{E}(\text{G})\textup{I}_{k}$ is a positive definite matrix under Assumption \ref{Ass2}. Therefore, the first $k$ eigenvectors corresponding to the first $k$ eigenvalues of $\mathrm{E}(\bm{Z}\bm{Z}^\top )$ are the basis of $\mathrm{Span}(\textup{B})$ and
$\Sigma_{X}^{-1}\mathrm{E}(Y^{2}\bm{X}\bm{X}^\top )\Sigma_{X}^{-1}-\Sigma_{X}^{-1}\mathrm{E}\left(\text{G}\right)\textup{I}_{p}$ is contained in the linear subspace Spanned by the basis matrix $\text{B}$.

When $\bm{X}\sim \mathrm{N}_{p}(\mu,\Sigma)$, 
without loss of generality, 
we can obtain   $\mathrm{E}( \bm{W}_{(2)}\bm{W}_{(2)}^\top \mid \bm{W}_{(1)}) = \text{I}_{p-k}$ by transforming $\bm{X}^{*}=\Sigma^{-1/2}\bm{X}$, then 
\begin{align*}
    \mathrm{E}&\Big[\mathrm{E}\left \{H\left(w_{1},\dots,w_{k}\right)\cdot w_{k+1}^{2}\mid \bm{W}_{(1)} \right \}\Big]\\&=\mathrm{E}\Big\{ H\left(w_{1},\dots,w_{k}\right) \mathrm{E}\left ( w_{k+1}^{2}\mid \bm{W}_{(1)} \right )\Big\}=\mathrm{E}\left(Y^{2}\right).
\end{align*}
Furthermore, let $\tilde{Y}^{2}=Y^{2}-\mathrm{E}(Y^{2})$, then $\mathrm{E}(\tilde{Y}^{2})=0$ and
\begin{equation*}
\Sigma^{-1}\mathrm{E}(\tilde{Y}^{2}\bm{X}\bm{X}^\top )\Sigma^{-1}= \text{B}\Gamma _{1}\text{B}^\top \equiv \text{B}\tilde{\textup{M}}.
\end{equation*}
This means that the non-zero $k$ eigenvectors corresponding to the non-zero $k$ eigenvalues of $\Sigma^{-1}\mathrm{E}(\tilde{Y}^{2}\bm{X}\bm{X}^\top )\Sigma^{-1}$ are the basis of $\mathrm{Span}(\text{B})$. 

\subsection{Proof of Theorem \ref{Thm2}}

By the Law of Large Number, we have
\begin{equation*}
	\dfrac{1}{n}\sum_{i=1}^{n}Y_{i}(\bm{X}_{i}-\bar{\bm{X}})=\dfrac{1}{n}\sum_{i=1}^{n}Y_{i}(\bm{X}_{i}-\mu)+\bar{Y}(\mu-\bar{\bm{X}})\stackrel{\mathrm{Pr}}{\longrightarrow} \mathrm{E}(Y\bm{X})=\Sigma_{X}\text{B}\Lambda.
\end{equation*}
By Lemma \ref{Lemma1} and $
	S_{n}\stackrel{\mathrm{Pr}}{\longrightarrow}\Sigma_{X}$, we have
\begin{equation*}
	S_{n}^{-1}\dfrac{1}{n}\sum_{i=1}^{n}Y_{i}(\bm{X}_{i}-\bar{\bm{X}})\stackrel{\mathrm{Pr}}{\longrightarrow}\Sigma_{X}^{-1}\Sigma_{X}\text{B}\Lambda=\text{B}\Lambda.
\end{equation*}
Let
\begin{equation*}
\bm{Z}_{i}=S_{n}^{-1}\left\{Y_{i}(\bm{X}_{i}-\bar{\bm{X}})\right\},
\end{equation*}
then,
\begin{equation*}
\hat{\bm{Z}}=\dfrac{1}{n}\sum_{i=1}^{n}\bm{Z}_{i}\stackrel{\mathrm{Pr}}{\longrightarrow}\text{B}\Lambda
%\ with\ the\ rate \ n^{1/2},
\end{equation*}
with convergence rate of $n^{1/2}$, and 
\begin{equation*}
\hat{\mathcal{Z}}=\dfrac{1}{n}\sum_{i=1}^{n}\bm{Z}_{i}\bm{Z}_{i}^\top \stackrel{\mathrm{Pr}}{\longrightarrow}\mathrm{E}\left(\bm{Z}\bm{Z}^\top \right)
\end{equation*}
with convergence rate $n^{1/2}$. 

From Theorem \ref{thm1}, the first $k$ eigenvectors corresponding to the first $k$ eigenvalues of $\mathrm{E}(\bm{Z}\bm{Z}^\top )$ are the basis of $\mathrm{Span}(\text{B})$. Consequently, the first $k$ eigenvectors corresponding to the first $k$ eigenvalues of $\hat{\mathcal{Z}}$, $\hat{\beta}_{1}, \ldots,\hat{\beta}_{k}$, converge to the corresponding rotational basis for $\mathrm{Span}(\text{B})$ with convergence rate of $n^{1/2}$.

Since the elements of $\mathrm{vec}(\hat{\mathcal{Z}})$ are moment estimators of the elements of $\bm{Z}\bm{Z}^\top $, by the central limit theorem,  we have 
\begin{equation*}
\sqrt{n}\left [ \mathrm{vec}\left(\hat{\mathcal{Z}}\right)- \mathrm{vec}\left\{\mathrm{E}\left(\bm{Z}\bm{Z}^\top \right)\right\} \right ]
\end{equation*}
converges in distribution to a multivariate normal random vector with mean vector $0$ and variance-covariance matrix $\mathrm{Var}\{ \mathrm{vec}(\bm{Z}\bm{Z}^\top ) \}$. Here, we derive the specific form of $\mathrm{Var}\{ \mathrm{vec}(\bm{Z}\bm{Z}^\top ) \}$. Using the techniques in the proof of Theorem \ref{thm1} in Appendix $A.2$, we can obtain the fourth conditional moment of $Y$ given $\bm{X}$ as
\begin{equation*}
\mathrm{E}\left(Y^{4}\mid \bm{X}\right)=U\left(\beta_{1}^\top \bm{X},\dots,\beta_{k}^\top \bm{X}\right). 
\end{equation*}
Given $\mathrm{E}(Y^{2}\bm{X}\bm{X}^\top )$, the essence of obtaining the variance-covariance matrix is calculating the fourth moment. For notation convenience, we use the formation of the Kronecker product, which contains the fourth moments. Following the definition of $\bm{W}$ and the result in the proof of Theorem \ref{thm1}, we have 
\begin{equation}
\begin{aligned}
    &\mathrm{Var}\left \{ \mathrm{vec}\left(\bm{Z}\bm{Z}^\top \right) \right \}=\mathrm{Var}\left \{ Y^{2}\mathrm{vec}\left(\bm{X}\bm{X}^\top \right) \right \}\\
    &=\mathrm{E}\left \{ Y^{2}\mathrm{vec}\left(\bm{X}\bm{X}^\top \right)\mathrm{vec}\left(\bm{X}\bm{X}^\top \right)^\top  Y^{2}\right \}\\
    & \qquad \qquad -\mathrm{E}\left \{ Y^{2}\mathrm{vec}\left(\bm{X}\bm{X}^\top \right)\right \}\times\mathrm{E}\left \{ Y^{2}\mathrm{vec}\left(\bm{X}\bm{X}^\top \right)\right \}^\top \\
    &=\mathrm{E}\left[ \mathrm{E}\left \{ Y^{4}\mathrm{vec}\left(\bm{X}\bm{X}^\top \right)\mathrm{vec}\left(\bm{X}\bm{X}^\top \right)^\top \mid \bm{X}\right \} \right]\\
    & \qquad \qquad-\mathrm{E}\Big[ \mathrm{E}\left \{ Y^{2}\mathrm{vec}\left(\bm{X}\bm{X}^\top \right)\mid \bm{X}\right \} \Big]\times\mathrm{E}\Big[ \mathrm{E}\left \{ Y^{2}\mathrm{vec}\left(\bm{X}\bm{X}^\top \right)\mid \bm{X}\right \} \Big]^\top \\
    &\propto \mathrm{E}\Big\{ \left(\bm{X}\bm{X}^\top \right)\otimes\left(\bm{X}\bm{X}^\top \right) U\left(\beta_{1}^\top \bm{X},\dots,\beta_{k}^\top \bm{X}\right)  \Big\}\\
    & \qquad \qquad-\mathrm{vec}\Big\{\mathrm{E}\left(Y^{2}\bm{X}\bm{X}^\top \right)\Big\}\times \mathrm{vec}\Big\{\mathrm{E}\left(Y^{2}\bm{X}\bm{X}^\top \right)\Big\}^\top ,
\end{aligned}
\end{equation}
where $\propto$ denotes the dimensional inequality and $\otimes$ denotes the Kronecker product. As the Kronecker product can incorporate all the elements of the variance-covariance matrix of the random matrix $\bm{Z}\bm{Z}^\top $,  it does not affect the convergence of elements even though the dimensions are different. 

We can obtain $\mathrm{E}\left(Y^{2}\bm{X}\bm{X}^\top \right)$ from the proof of Theorem \ref{thm1} in Appendix $\textup{A.2}$. Then,  we require the following: 
\begin{align*}
 \mathrm{E}&\left\{ Y^{4}\left(\bm{X}\bm{X}^\top \right) \otimes\left(\bm{X}\bm{X}^\top \right)\right\}\\
 &=\mathrm{E}\Big\{ \left(\bm{X}\bm{X}^\top \right) \otimes\left(\bm{X}\bm{X}^\top \right) U\left(\beta_{1}^\top \bm{X},\dots,\beta_{k}^\top \bm{X}\right)  \Big\}\\
 &=\mathrm{E}\Big\{ \left(\text{C}^\top \bm{W}\bm{W}^\top \text{C}\right)\otimes\left(\text{C}^\top \bm{W}\bm{W}^\top \text{C}\right) U\left(\beta_{1}^\top \bm{X},\dots,\beta_{k}^\top \bm{X}\right)  \Big\}\\
 &=\mathrm{E}\Big[ U\left(w_{1},\dots,w_{k}\right)\mathrm{E}\Big\{ \left(\text{C}^\top \bm{W}\bm{W}^\top \text{C}\right)\otimes\left(\text{C}^\top \bm{W}\bm{W}^\top \text{C}\right)\mid \bm{W}_{(1)} \Big\} \Big]\\
 &=\mathrm{E}\Big\{ U\left(w_{1},\dots,w_{k}\right)\left(\text{C}^\top \otimes \text{C}^\top \right)\mathrm{E}\left(\bm{W}\bm{W}^\top \otimes \bm{W}\bm{W}^\top \mid \bm{W}_{(1)}\right)\left(\text{C}\otimes \text{C}\right) \Big\}\\
 &=\left(\text{C}^\top \otimes \text{C}^\top \right)\mathrm{E}\Big\{ U\left(w_{1},\dots,w_{k}\right)\mathrm{E}\left(\bm{W}\bm{W}^\top \otimes \bm{W}\bm{W}^\top \mid \bm{W}_{(1)}\right) \Big\}\left(\text{C}\otimes \text{C}\right).
\end{align*}
As a result, we have $\mathrm{E}\left( w_{i}w_{j}w_{s}w_{t}\mid \bm{W}_{(1)} \right)$, where $i,j,s,t\in \{1,\dots,p\}$. Considering $\Sigma_{X}=\textup{I}_{p}$, the form of the elements in $\mathrm{E}\left(\bm{W}\bm{W}^\top  \otimes \bm{W}\bm{W}^\top \mid \bm{W}_{(1)}\right)$ can be expressed as 
\begin{align*}
\mathrm{E}\left ( w_{i}w_{j}w_{s}w_{t}\mid \bm{W}_{(1)} \right )=\left\{\begin{matrix}
 w_{i}w_{j}w_{s}w_{t}, & i,j,s,t \in \{1,\dots,k\},\\ 
w_{i}w_{j}\mathrm{E}\left ( w_{s}w_{t}\mid \bm{W}_{(1)} \right ), & i,j \in \{1,\dots,k\}, s=t \in \{k+1,\dots,p\},\\
w_{i}\mathrm{E}\left ( w_{j}w_{s}w_{t}\mid \bm{W}_{(1)} \right ), & i \in \{1,\dots,k\},
j=s=t \in \{k+1,\dots,p\},\\
\mathrm{E}\left ( w_{i}w_{j}w_{s}w_{t}\mid \bm{W}_{(1)} \right ), & i=j=s=t \in \{k+1,\dots,p\},\\ 
0, & \text{otherwise}.
\end{matrix}\right.
\end{align*}
Hence, $\mathrm{E}\left\{ Y^{4}\left(\bm{X}\bm{X}^\top \right) \otimes\left(\bm{X}\bm{X}^\top \right)\right\}$, and the same technique can be applied when $\Sigma_{X}=\Sigma$.

\subsection{R code for the proposed PSRFR}

{\tt

\noindent PSRFR <- function(X, y, r) \{

  n <- nrow(X)          \#\# Number of observations
  
  dim <- ncol(X)        \#\# Dimensionality of X
  
  \#\# Create data matrix combining X and y
  
  data <- cbind(X, t(y))
  
  \#\# Mean calculations
  
  my <- mean(y)
  
  mx <- colMeans(X)     \#\# Vector of means for each column in X
  
  \#\# Centering matrix X
  
  mxx <- matrix(mx, nrow = n, ncol = dim, byrow = TRUE)
  
  z <- (X - mxx) * data[, dim + 1]  \#\# Computing z
  
  \#\# Covariance of X and related calculations
  
  sigmax <- \text{Cov}(X)
  
  invSigmax <- solve(sigmax)
  
  K <- invSigmax \%*\% (t(z) \%*\% z / n) \%*\% invSigmax
  
 \#\# Eigenvalue decomposition
  
  Kv <- eigen(K)
  
  \#\# Return a list of eigenvalues and eigenvectors
  
  return(list(values = Kv\$values, vectors = Kv\$vectors[,1:r]))
  
\noindent \}
}

\newpage

\newpage

\bibliographystyle{apalike}
\bibliography{refs.bib}
\end{document}